\documentclass[english]{emulateapj2}
\usepackage[T1]{fontenc}
\usepackage[latin9]{inputenc}
\usepackage{verbatim}
\usepackage{amssymb}
\usepackage{pdflscape}

\newcommand{\Mp}{M_{\rm p}}

\newcommand{\apla}{a_{\rm p}}
\newcommand{\ep}{e_{\rm p}}
\newcommand{\np}{n_{\rm p}}
\newcommand{\Mc}{M_{\rm c}}
\newcommand{\ac}{a_{\rm c}}
\newcommand{\ec}{e_{\rm c}}
\newcommand{\nc}{n_{\rm c}}
\begin{document}

\title{Tidal Evolution of Close-in Planets}

\author{Soko Matsumura\altaffilmark{1,2}}
\author{Stanton J.\ Peale\altaffilmark{3}}
\author{Frederic A.\ Rasio\altaffilmark{1,4}}

\altaffiltext{1}{Department of Physics and Astronomy, Northwestern
University, Evanston, IL 60208.}

\altaffiltext{2}{Currently Center for Theory and Computation Fellow,
University of Maryland,
soko@astro.umd.edu.}

\altaffiltext{3}{Physics Department, University of California, Santa
Barbara, CA 93106.}

\altaffiltext{4}{Also Center for Interdisciplinary Exploration and
Research in Astrophysics
  (CIERA), Northwestern University.}

\begin{abstract}
Recent discoveries of several transiting planets with clearly
non-zero eccentricities and some large obliquities started changing
the simple picture of close-in planets having circular and
well-aligned orbits.
Two major scenarios to form such close-in planets are planet
migration in a disk, and planet--planet interactions combined with
tidal dissipation. The former scenario can naturally produce a
circular and low-obliquity orbit, while the latter implicitly
assumes an initially highly eccentric and possibly high-obliquity
orbit, which are then circularized and aligned via tidal
dissipation.

Most of these close-in planets experience orbital decay all the way
to the Roche limit as the previous studies showed.
We investigate the tidal evolution of transiting planets on
eccentric orbits, and find that there are two characteristic
evolution paths for them, depending on the relative efficiency of
tidal dissipation inside the star and the planet.
Our study shows that each of these paths may correspond to
migration, and scattering scenarios, respectively. We further point
out that the current observations may be consistent with the
scattering scenario, where the circularization of an initially 
eccentric orbit occurs 
before the orbital decay primarily due to tidal
dissipation in the planet, while the alignment of the stellar spin
and orbit normal occurs on the similar timescale to the orbital
decay largely due to dissipation in the star.
We also find that even when the stellar spin-orbit misalignment is
observed to be small at present, some systems could have had a
highly misaligned orbit in the past, if their evolution is dominated
by tidal dissipation in the star.

Finally, we also re-examine the recent claim by Levrard et.~al. that
all orbital and spin parameters, including eccentricity and stellar
obliquity, evolve on a similar timescale to orbital decay.
This counter-intuitive result turns out to have been caused by a typo 
in their numerical code.
Solving the correct set of tidal equations, we find that the
eccentricity behaves as expected, with orbits usually circularizing
rapidly compared to the orbital decay rate.
\end{abstract}


\section{Introduction}

More than 450 exoplanets have been discovered so far. Out of about
360 extrasolar planetary systems, roughly $30\%$ possess close-in
planets with semimajor axis $a \lesssim0.1\,$AU. Also, there are 18
out of 45 multiple-planet systems with at least one close-in planet.
The mean eccentricity for extrasolar planets with $a <0.1\,$AU is
close to zero, while for planets beyond 0.1 AU, it is $e\simeq
0.25$. This sharp decline in eccentricity close to the central star
is usually explained as a result of efficient eccentricity damping
due to tidal interactions between the star and the planet
\citep[e.g.,][]{Rasio96,Jackson08}. Additionally, there are currently at least
26 systems with measurements of the projected stellar obliquity
angle $\lambda$ (see Table~\ref{tab1}) through the
Rossiter-McLaughlin (RM) effect
\citep{Rossiter24,McLaughlin24,Ohta05,Gaudi07}. Although many
systems have projected stellar obliquities consistent with zero
within $2\sigma$ \citep[e.g.,][]{Fabrycky09}, suggesting
near-perfect spin-orbit alignment, there are now several planetary
systems that are clearly misaligned \citep{Triaud10ap}. Examples
include XO-3, HD~80606, and WASP-14, which are in prograde orbits
with $\lambda\simeq 37.3 \pm 3.7$, $53^{+34}_{-21}$, and $-33.1\pm
7.4\,$degrees, respectively
\citep{Winn09XO3,Winn09HD80606,Johnson09}, as well as HAT-P-7,
WASP-2, WASP-8, WASP-15, and WASP-17, which have {\em retrograde}
orbits with $\lambda\simeq 182.5\pm9.4$, $-153^{+15}_{-11}$,
$-120\pm4$, $-139.6^{+4.3}_{-5.2}$, and
$-147.3^{+5.5}_{-5.9}\,$degrees, respectively
\citep{Winn09HATP7,Triaud10ap}.

The standard planet formation theory predicts that giant planets are
formed beyond the so-called ``ice line'' ($a\gtrsim3\,$AU), where
solid material is abundant due to condensation of ice. To explain
the orbital properties of close-in planets, two different scenarios
have been proposed. Both of them could potentially explain the
proximity of these planets to the stars, but they predict different
distributions for orbital inclinations, and possibly eccentricities.
One scenario is orbital migration of planets due to gravitational
interactions with gas or planetesimal disks
\citep[e.g.,][]{Goldreich80,Ward97,Murray98}, which would naturally
bring planets inward from their formation sites. The scenario can
also account for the observed small eccentricity and obliquity seen
in the majority of close-in planets, because the disks tend to damp
eccentricity and inclination of the planetary orbit
\citep[e.g.,][]{Goldreich80,Papaloizou00}.
Alternatively, such close-in planets can be formed by tidally
circularizing a highly eccentric orbit. A natural way of initially
increasing the orbital eccentricity is through gravitational
interactions between several planets. Although such interactions
alone may not be able to populate the inner region of planetary
systems \citep[less than $0.1-1\,$AU,
e.g.,][]{Adams03,Chatterjee08,Matsumura10}, the orbital eccentricity
can be increased due to scattering, ejection, or Kozai cycles
\citep{Kozai62,Lidov62}, so that the pericenter of the planetary
orbit becomes small enough for tidal interactions with the central
star to become important \citep[e.g.,][]{Nagasawa08}. These
gravitational interactions also tend to increase the orbital
inclination. \citet{Chatterjee08} performed a number of dynamical
simulations of three-planet systems, and showed that the final mean
inclination of planetary orbits is about 20 degrees, and that some
planets could end up on retrograde orbits (about $2\%$, S.\
Chatterjee, private communication).
When Kozai cycles increase the orbital eccentricity, the
process is called Kozai migration \citep{Wu03}. Kozai migration
occurring in binary systems may be responsible for at least some of
the close-in planets \citep{Fabrycky07,Wu07,Triaud10ap}. One of the
goals of our study is to explore the possibility and implications of
forming close-in planets via tidal circularization of a highly
eccentric planet.

Independent of their formation mechanism, these close-in planets are
currently subject to strong tidal interactions with the central
star, and such interactions could dominate the orbital evolution of
these planets. In multi-planet systems, secular planet--planet
interactions may also affect the orbital evolution. However, in
Section~2, we will show that it is unlikely that the current and
future evolution of the observed close-in planets is strongly
affected by any known or yet-to-be-detected companion \citep[see
also][]{Matsumura08}. In this paper, we investigate tidal evolution
of close-in planets to distinguish the two formation scenarios of
them.

Tidal evolution in a two-body system leads to either a stable
equilibrium state, or to orbital decay all the way to the Roche
limit
\citep[``Darwin instability''][]{Darwin1879}. Such a study for
exoplanetary systems was first done by \cite{Rasio96}, who suggested
that 51~Peg~b, the only close-in planet known at the time, would be
Darwin unstable. In a recent paper, \citet[hereafter,
LWC09]{Levrard09} investigated the tidal evolution of all transiting
planets, and pointed out that most of these planets (except
HAT-P-2b) are indeed Darwin unstable, and thus undergo continual
orbital decay, rather than arrive at a stable, equilibrium orbit
(see Section~4.1, and Table~\ref{tab2} for updated results). These
Darwin-unstable planets may eventually be accreted by the central
star, which has been suggested for some systems observationally
\citep[e.g.][]{Gonzalez97,Ecuvillon06} and numerically
\citep{Jackson09ap}.

There is some confusion in the literature concerning the evolution
timescales for the various orbital elements. LWC09 studied tidal
evolution of transiting planets by taking into account energy
dissipation in both the star and the planet. They concluded that all
orbital and spin parameters, with the exception of the planetary
spin, would evolve on a timescale comparable to the orbital decay
timescale. This would imply that both circularity of the orbits and
spin-orbit alignment seen in many systems are primordial, because
neither obliquity nor eccentricity could be damped to zero before
complete spiral-in and destruction of the planet. Unfortunately,
after much investigation and comparison with their work, we
determined that LWC09 had a typo in their code (confirmed by B.\
Levrard, personal communication), which made them underestimate the
energy dissipation inside the planet by several orders of magnitude.
By integrating the correct set of tidal evolution equations, we find
that there are two characteristic evolutionary paths depending on
the relative efficiency of tidal dissipation inside the star and the
planet. When the dissipation in the planet dominates, the
eccentricity damping time is shorter than the orbital decay time
($\tau_{e}\ll \tau_{a}$; see Section~4 for details), exactly as
expected intuitively \citep[e.g.,
][]{Rasio96,DobbsDixon04,Mardling04}. On the other hand, when the
dissipation in the star dominates, the eccentricity damps on a
similar timescale to the orbital decay. We will show that the latter
path is fundamentally different from what is suggested by LWC09 in
Section~4.2.

There have been many other recent studies of tidal evolution for
exoplanets. \cite{Jackson08} emphasized the importance of solving
the coupled evolution equations for eccentricity and semi-major
axis. Integrating their tidal evolution equations backwards in time,
they showed that the ``initial'' eccentricity distribution of
close-in planets matches more closely that of planets on wider
orbits; they suggested that gas disk migration is therefore not
responsible for all the close-in planets. We reconsider the validity
and implication of such a study in Section~5. \cite{Barker09}
studied the evolution of close-in planets on inclined orbits, by
including the effect of magnetic braking \citep{DobbsDixon04}, and
pointed out that a true tidal equilibrium state is never reached in
reality, since the total angular momentum is not conserved due to
magnetized stellar winds. They also showed that neglecting this
effect could result in a very different predicted evolution for the
systems they considered. Throughout this paper, we compare the tidal
evolutions with and without the effect of magnetic braking. Another
potentially important effect caused by tidal dissipation is the
inflation of the planetary radius \citep{Bodenheimer01,Gu03}. Many
groups have explored this possibility, motivated by observations of
inflated radii for some transiting planets
\citep[e.g.][]{Barge08,Alonso08,JohnsKrull08,Gillon09a}. It appears
that at least some of these inflated planets could be explained as a
result of past tidal heating \citep{Jackson08,Miller09,Ibgui09}.
More recently, \cite{Leconte10} revisited this problem, and pointed
out that the truncated tidal equations used in many previous studies
could lead to an erroneous tidal evolution for moderate to high
eccentricity ($e\gtrsim 0.2$).  Solving the complete set of tidal
equations, they showed that the orbital circularization occurs much
earlier than previously estimated, and thus only moderately bloated
hot Jupiters could be explained as a result of tidal heating. In
this paper we neglect this effect entirely and treat the planetary
radius as constant for simplicity (but see Section~5.1).

The outline of the paper is as follows. First, we justify our
approach in Section~2 by showing that the current/future evolution
of these planets is likely dominated by tidal dissipation, rather
than by their gravitational interaction with a more distant object
(planet on a wider orbit or distant binary companion). Then we
present our set of tidal evolution equations in Section~3. We
re-examine the tidal stability of transiting planets, and
investigate the tidal evolution forward in time in Section~4. We
identify two characteristic evolution paths for Darwin-unstable
planets, and also discuss the results of LWC09. In Section~5 we
explore the past evolution of transiting planets and its
implication. Our results imply that each evolutionary path may be
consistent with migration, and gravitational-interaction-induced
formation scenarios of transiting planets, respectively. We also
point out that in a limited case, it's possible to have a
significant stellar obliquity damping before the substantial orbital
decay to the Roche limit. In Section~6, we study the different
definitions of tidal quality factors, and investigate their effects
on evolution.
Finally, in Section~7, we discuss and summarize our results.

\section{Possible Companions to Close-in Planets}
In this section, we assess the importance of a known/unknown
companion on the current and future evolution of close-in transiting
planets. Secular or resonant interactions with other planets or
stellar companions could potentially perturb the planetary orbits
significantly. For example, when there is a large mutual inclination
between their orbits ($\gtrsim39.2^{\circ}$), Kozai-type
(quadrupole) perturbations can become important
\citep{Kozai62,Lidov62}. Such highly misaligned systems may
naturally occur for binary systems with semimajor axis
$\gtrsim30-40\,$AU \citep{Hale94}, or as a result of planet--planet
scattering \citep{Chatterjee08,Nagasawa08}. For smaller mutual
inclinations, octupole-level perturbations may still moderately
excite the orbital eccentricity, as long as the companion has a
non-circular orbit.

For this secular perturbation from a companion to be affecting the
current and future evolution of close-in planets, it must occur fast
compared to other perturbations that cause orbital precession. These
competing perturbations include general relativistic (GR)
precession, tides, as well as rotational distortions
\citep{Holman97,Sterne39}.  Since the effects of pericenter
precession due to stellar and planetary oblateness are usually small
compared to those caused by GR precession
\citep{Kiseleva98,Fabrycky07}, we simply neglect rotational
distortions here.

The pericenter precession timescales corresponding to GR,
Kozai-type perturbations (due to a high-inclination
perturber), and secular coupling to a low-inclination perturber, can
be written as follows \citep{Kiseleva98,Fabrycky07,Zhou03,Takeda08}:
\begin{eqnarray}
\tau_{\rm GR}&=& \frac{2 \pi c^2 \apla}{3G(M_*+\Mp)\np}(1-\ep^2)
\label{tgr} \\
\tau_{\rm Kozai}&=& \frac{4\np}{3 \nc^2} \left(\frac{M_*+\Mp+\Mc}
{\Mc}\right)(1-\ec^2)^{3/2} \label{tkozai} \\
\tau_{\rm pp}&=& \frac{4\pi}{\left(c_1+c_2\right) \pm
\sqrt{\left(c_1-c_2\right)^2+4c_0^2c_1c_2}} \ ,
\end{eqnarray}
where the subscript $*$, p and c indicate the central star, the
planet, and the companion body, respectively, while $c$ is the speed
of light. In $\tau_{\rm pp}$,
\begin{eqnarray}
c_0&=&b^{(2)}_{3/2}\left(\frac{\apla}{\ac}\right)/b^{(1)}_{3/2}\left(\frac{\apla}{\ac}\right), \\
c_1&=&\frac{1}{4}\np\frac{\Mc}{M_*+\Mp}\left(\frac{\apla}{\ac}\right)^2
b^{(1)}_{3/2}\left(\frac{\apla}{\ac}\right), \\
c_2&=&\frac{1}{4}\nc\frac{\Mp}{M_*+\Mc}\left(\frac{\apla}{\ac}\right)
b^{(1)}_{3/2}\left(\frac{\apla}{\ac}\right),
\end{eqnarray}
with the standard Laplace coefficients $b^{(i)}_{3/2}(\apla/\ac)$
($i=1,2$). For the secular timescale, the upper sign is chosen when
$\Mp<\Mc$ and the lower one is chosen when $\Mp>\Mc$.  Such an
approximation is reasonable when the system is hierarchical
\citep{Takeda08} and thus
\[
\frac{a_p/a_c}{1-3\left(\Mc/\Mp\right)\sqrt{a_p/a_c}/b^{(1)}_{3/2}} \ll 1 \ .
\]
Following the approach of \cite{Matsumura08}, we compare these
various precession timescales to determine which perturbation
dominates.

\begin{figure}
\plottwo{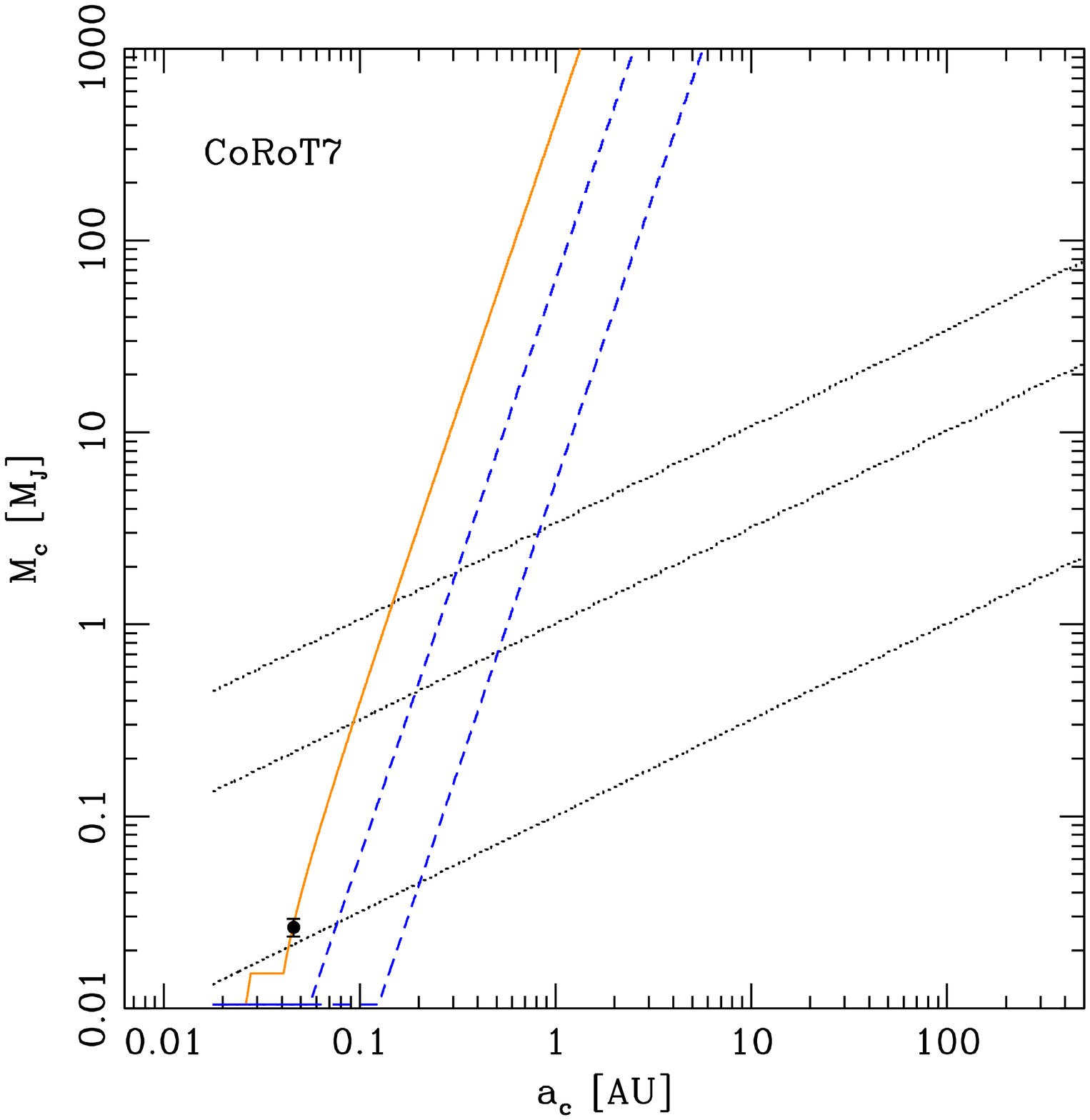}{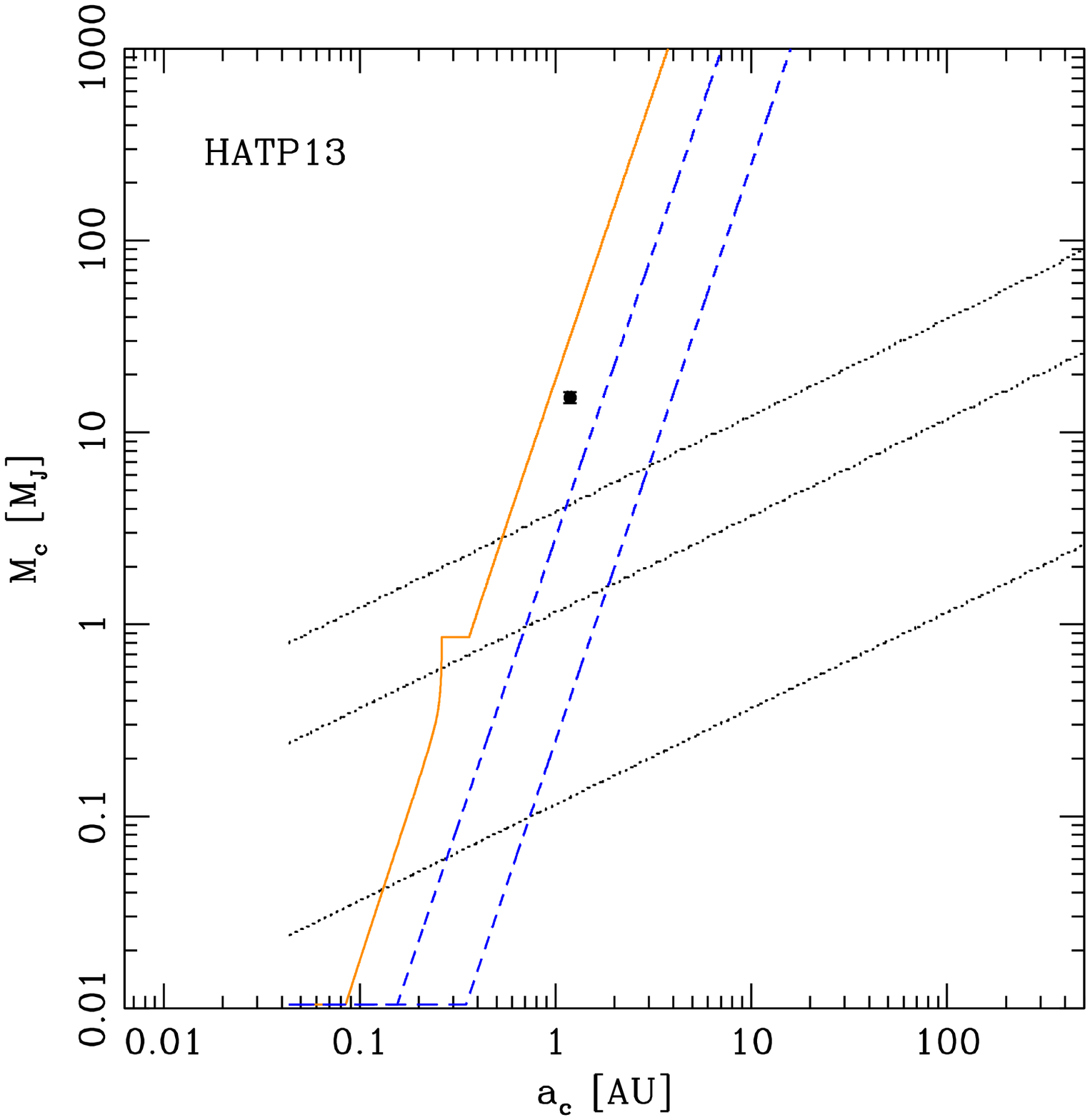}

\caption{Mass and semi-major axis of a hypothetical companion which
can affect the orbital evolution of the observed close-in planets in
CoRoT-7 and HAT-P-13. Orange line indicates the region within which
the secular timescale becomes short compared to the GR timescale.
The kink occurs where the approximation for the secular timescale
breaks down (see text). Left and right blue dashed lines indicate
similar boundaries for Kozai cycles with the companion
eccentricity of 0.2, and 0.9, respectively. Black dotted lines
indicate the radial velocity detection limits for 3, 30, and
100\,m/s. Both of these systems have a known second planet, which is
indicated by a filled circle. \label{fig0}}

\end{figure}

Fig.~\ref{fig0} shows the resulting constraints on the mass and
orbital radius of the hypothetical planetary/stellar companion for
CoRoT-7b and HAT-P-13b.   These are the only two known multi-planet
systems with a transiting planet, and the companions are shown as
the filled circles in the plot. In order to perturb the inner planet
significantly despite the GR precession, the companion must exist
left of the blue lines to induce Kozai oscillations, and left
of the orange line for octupole perturbations to be important. The
kink seen in the orange line occurs where $\Mp\sim\Mc$ and thus our
approximation in calculating the secular timescale breaks down. For
HAT-P-13, we find that HAT-P-13c is located left of the blue lines,
but right of the orange one. Thus, we expect that HAT-P-13c can
perturb the orbit of HAT-P-13b significantly only if they have a
large mutual inclination of $\gtrsim39.2^{\circ}$. For CoRoT-7, on
the other hand, we find that the secular timescale due to CoRoT-7c
is comparable to the GR timescale ($\tau_{\rm pp}\sim\tau_{\rm
GR}$). As shown by \cite{Ford00} \citep[see also][]{Adams06}, such a
resonance can increase the eccentricity significantly. However, for
small eccentricities $e\sim0.001$, we do not find any significant
eccentricity oscillation below a mutual inclination of about
$40^{\circ}$. Thus it is possible that the eccentricity of CoRoT-7b
is significantly oscillating if the misalignment of the orbit of
CoRoT-7c with respect to CoRoT-7b is large.

\begin{figure}
\plotone{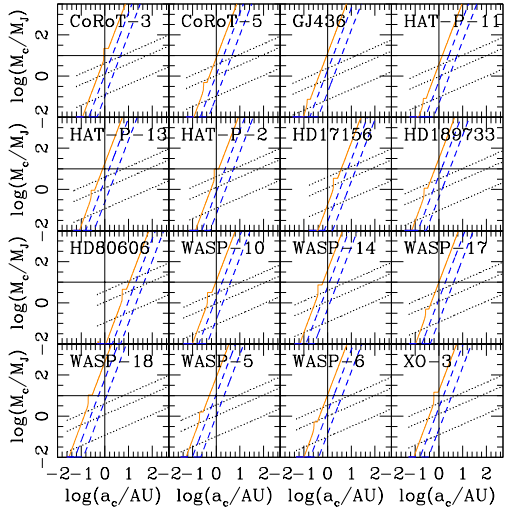}

\caption{Similar plots to Fig.~\ref{fig0} for all transiting planets
with an eccentric orbit. Vertical and horizontal lines are drawn at
1 AU and $10M_J$ for comparison.  It is clear that all systems
except HD~80606 are unlikely to have a companion which can cause
significant secular perturbations (faster than the GR precession).
\label{companionall} }

\end{figure}

In Fig.~\ref{companionall}, we present similar plots for all
transiting planets on an eccentric orbit. Vertical and horizontal
lines are drawn for reference at $1\,$AU and $10M_J$, respectively.
It is clear that all systems except HD~80606 are unlikely to have a
companion which has a secular perturbation timescale shorter than
the GR timescale (left of orange and blue lines), and at the same
time does not cause detectable radial-velocity variations (below
black dotted lines, or beyond the observation limit in semimajor
axis). The figure predicts that, if a hypothetical planet which can
cause a significant secular perturbation is too small to be observed
(i.e.,~a planet exists left of the orange line and below black
dotted lines), its mass would be comparable to Earth or smaller. The
figure also predicts that, if such a planet cannot be observed now
simply because we are not observing long enough (i.e.,~a planet
exists left of the orange line, and beyond, for example, $1\,$AU),
its mass would be comparable to brown dwarfs or larger.

HD~80606b is in a wide ($\sim 1200\,$AU) binary system
\citep{Eggenberger04}. However, as can be seen in the figure, this
companion cannot induce secular perturbations fast enough compared
to the GR precession.

Resonant interactions can work in a similar way, but it is unlikely
that all of these planets have such a companion. Thus, the evolution
of currently observed transiting planets with an eccentric orbit is
likely dominated by tidal dissipation rather than interactions with
outer companions.

\section{Tidal Evolution Equations}

In this paper, we numerically study the evolution of observed
transiting planets by integrating a set of equations describing the
tidal interactions, and by assuming that the known/unknown
companions' effects can be neglected.

We follow the general approach of the equilibrium tide model with
the weak friction approximation \citep{Darwin1879}. The effects of
the dynamical tide are neglected for simplicity. In absence of any
dissipation, the tidally distorted body is assumed to take the
equilibrium shape that adjusts itself to the external potential
field of the tide-raising body. When the dissipation is
non-negligible, the equilibrium surface is either lagged or led,
depending on whether the spin frequency is smaller or larger than
the orbital frequency. In the limit of small viscosities, this phase
lag ($\phi$) can be approximated to be proportional to the tidal
forcing frequency ($\sigma$) as $\phi\sim \Delta t \sigma$.
This allows us to interpret the phase lag as the tidal bulge that
could have been raised a constant time $\Delta t$ ago in an inviscid
case. Within the context of this model, we can derive the secular
equations which are valid for any value of eccentricity and
obliquity, by following the approach of \cite{Alexander73,Hut81}
\citep[also see][]{Leconte10}. Taking into account tides raised both
on the central star by the planet and on the planet by the star, the
complete set of tidal equations can be written as follows for the
semi-major axis $a$, eccentricity $e$, stellar obliquity
$\epsilon_*$, planetary spin $\omega_{p}$, and stellar spin
$\omega_{*}$, respectively \citep[e.g.,][]{Hut81,Levrard07}. Note
that we assume that the planetary obliquity is zero (i.e., the
equatorial plane of the planet coincides with the orbital plane). We
explicitly write down all equations in order to compare our results
with LWC09:
%
\begin{eqnarray}
\frac{da}{dt} & = & 6k_{2,*}\Delta t_*n\frac{M_{p}}{M_{*}}\frac{R_{*}^{5}}{a^{4}}
\frac{\left[(1-e^{2})^{3/2}f_{2}(e^{2})\omega_{*}\cos\epsilon_*-f_{1}(e^{2})n \right]}{(1-e^{2})^{15/2}} \label{eq1} \\
 & + & 6k_{2,p}\Delta t_pn\frac{M_{*}}{M_{p}}\frac{R_{p}^{5}}{a^{4}}
\frac{\left[(1-e^{2})^{3/2}f_{2}(e^{2})\omega_{p}-f_{1}(e^{2})n \right]}{(1-e^{2})^{15/2}} \nonumber\\
\frac{de}{dt} & = & 27k_{2,*}\Delta t_*n\frac{M_{p}}{M_{*}}\frac{R_{*}^{5}}{a^{5}}
\frac{\left[\frac{11}{18}(1-e^{2})^{3/2}f_{4}(e^{2})\omega_{*}\cos\epsilon_*
-f_{3}(e^{2})n \right]}{e^{-1}(1-e^{2})^{13/2}} \label{eq2} \\
 & + & 27k_{2,p}\Delta t_pn\frac{M_{*}}{M_{p}}\frac{R_{p}^{5}}{a^{5}}
\frac{\left[\frac{11}{18}(1-e^{2})^{3/2}f_{4}(e^{2})\omega_{p}-f_{3}(e^{2})n \right]}{e^{-1}(1-e^{2})^{13/2}} \nonumber\\
\frac{d\epsilon_*}{dt} & =- & 3k_{2,*}\Delta t_*n \frac{M_{p}}{M_{*}+M_{p}}\frac{M_{p}}{M_{*}}\,
\frac{R_{*}^{3}}{a^{3}}\frac{n}{\omega_{*}}\frac{\sin\epsilon_*}{\alpha_{*}}  \nonumber \\ 
& \times &  \frac{\left[\, f_{2}(e^{2})n
-\frac{1}{2}(\cos\epsilon_*-\eta)(1-e^{2})^{3/2}f_{5}(e^{2})\omega_{*}\right]}{(1-e^{2})^{6}} 
\label{eq3} \\
\frac{d\omega_{p}}{dt} & = & 3k_{2,p}\Delta t_pn\frac{M_{*}}{M_{*}+M_{p}}\frac{M_{*}}{M_{p}}\,
\frac{R_{p}^{3}}{a^{3}}\frac{n}{\alpha_{p}}  \nonumber \\
& \times & \frac{\left[\, f_{2}(e^{2})n-(1-e^{2})^{3/2}f_{5}(e^{2})\omega_{p}\right]}{(1-e^{2})^{6}} \label{eq4}\\
\frac{d\omega_{*}}{dt} & = & 3k_{2,*}\Delta t_*n\frac{M_{p}}{M_{*}+M_{p}}\frac{M_{p}}{M_{*}}\,
\frac{R_{*}^{3}}{a^{3}}\frac{n}{\alpha_{*}} \nonumber \\
& \times & \frac{\left[\, f_{2}(e^{2})n\,\cos\epsilon_*
-\frac{1}{2}(1+\cos^{2}\epsilon_*)(1-e^{2})^{3/2}f_{5}(e^{2})\omega_{*}\right]}{(1-e^{2})^{6}} \label{eq5}  \ .
\end{eqnarray}

The subscripts $*$ and $p$ denote the star and planet, respectively.
These equations agree with those of \cite{Hut81} in the limit of
small $\epsilon_*$. In LWC09,
$\eta\equiv\alpha_{*}\frac{M_{*}+M_{p}}{M_{p}}\frac{R_{*}^{2}}{a^{2}}(1-e^{2})^{-1/2}\frac{\omega_{*}}{n}$
is set to zero (B.\ Levrard, private communication). This term
$\eta$ is smaller than $\cos\epsilon_*$ for most systems, but can be
comparable to, or larger than $\cos\epsilon_*$ for some systems.
Examples include CoRoT-1, HD~149026, Kepler-4, Kepler-8, and
WASP-17.
In the above equations, $k_{2}$ is the Love number for the
second-order harmonic potential \citep{Love44}, $\Delta t$ is a
constant time lag, $n$ is the mean motion, and $\alpha$ is the
square of the radius of gyration with $\alpha_{*}=0.06$, and
$\alpha_{p}=0.26$. The eccentricity functions $f_1(e^2)-f_5(e^2)$
are defined as follows as in \citet{Hut81}.
%
\begin{eqnarray*}
f_{1}(e^{2}) &=& 1+ \frac{31}{2}e^2 + \frac{255}{8}e^4 + \frac{185}{16}e^6 + \frac{25}{64}e^8 \\
f_{2}(e^{2}) &=& 1+ \frac{15}{2}e^2 + \frac{45}{8}e^4 + \frac{5}{16}e^6 \\
f_{3}(e^{2}) &=& 1+ \frac{15}{4}e^2 + \frac{15}{8}e^4 + \frac{5}{64}e^6 \\
f_{4}(e^{2}) &=& 1+ \frac{3}{2}e^2 + \frac{1}{8}e^4  \\
f_{5}(e^{2}) &=& 1+ 3e^2 + \frac{3}{8}e^4
\end{eqnarray*}

Although it is important to solve the coupled semimajor axis and
eccentricity evolution equations \citep{Jackson08}, the semi-major
axis evolution of a planetary system is likely dominated by the
energy dissipation in the star. This is because, in an eccentric
orbit, a gaseous planet's rotation will be tidally damped to an
asymptotic state that is somewhat faster than a value synchronous
with the orbital mean motion. The tidal torque is strongest at
pericenter where the orbital angular velocity exceeds the orbital
mean motion, and therefore the tidal torque averaged around the
orbit vanishes when the rotation rate exceeds the mean motion $n$.
This asymptotic state is often referred to as pseudo-synchronous
rotation. When the planetary spin approaches pseudo-synchronization
with the orbit normal $\omega_p\sim n$,
the contribution from the second term in Eq.~\ref{eq1} becomes
negligible for planets with small eccentricities. Therefore, unless
the eccentricity is very high, the orbital evolution is largely
determined by tidal dissipation in the star.

It's not immediately clear whether the tidal energy dissipation
leads to either orbital decay or orbital expansion. For all
transiting systems except CoRoT-3, CoRoT-6, HAT-P-2, HD~80606, and
WASP-7, the host star rotates slowly compared to the orbit (i.e.,
$\omega_{*}<n$, see Table~\ref{tab2}). Therefore, tidal dissipation
in the star tends to lead to orbital decay, by transferring angular
momentum from the orbit to the stellar spin. On the other hand, when
the host star is rapidly spinning, planets could migrate outward,
which may have prolonged the lifetime of some of the exoplanets
\citep{DobbsDixon04}. In Section~4.1, we show that HAT-P-2 is Darwin
unstable and migrating inward, while the other systems with a
rapidly rotating host star are evolving toward the stable tidal
equilibria.  More specifically, CoRoT-3, CoRoT-6, and WASP-7 are
currently migrating \emph{outward} toward the stable tidal
equilibria, while HD~80606 is migrating inward toward the stable
state.

By comparing Eq.~\ref{eq1} and \ref{eq2}, we see that the
eccentricity may be damped on a similar timescale to the semimajor
axis, when the eccentricity damping is dominated by tidal
dissipation in the star (i.e., the first term in Eq.~\ref{eq2} is
much larger than the second term). We discuss this further in the
next section.

In some of our calculations we also take into account the stellar
magnetic braking effects, and thus the loss of angular momentum due
to stellar winds. For this purpose, we assume that the Skumanich's
Law describes the decrease of the stellar spin sufficiently well so
that the average surface rotation velocities of stars that are not
interacting with close-in planets can be related to the stellar age
($\tau_{\rm age}$) as $V_*\sin i_*\propto 1/\sqrt{\tau_{\rm age}}$
\citep{Skumanich72}. From this relation, the change in stellar spin
can be written as follows.
\begin{equation}
\left(\frac{d\omega_{*}}{dt}\right)_{\rm mb}=\frac{\dot{V_*}}{R_*}\simeq -\frac{\gamma}{2}
\frac{R_*^2}{\tau_{\rm age,\,0}\,V_{*,0}^2}\,\omega_{*}^3 \equiv -\beta\,\omega_{*}^3 \,
\end{equation}
where $\gamma$ is a calibration factor, and the subscript $0$
denotes the normalization factors. By choosing $V_{*,0}=4\,$km/s and
$\tau_{\rm age,\,0}=1\,$Gyr as in \cite{DobbsDixon04}, we can define
$\beta\equiv\gamma1.5\times 10^{-14}\,$yr. We adopt $\gamma=0.1$ for
F stars, and $\gamma=1$ for G, K, and M stars as in \cite{Barker09}.
For F stars, the smaller calibration factor is chosen since magnetic
braking is less efficient due to the very thin, or completely absent
outer convective layer. We add this $-\beta\,\omega_*^3$ to
Eq.~\ref{eq5} when including the effects of magnetic braking.

\subsection{Tidal Quality Factors}
It is common to describe the dissipation inefficiency of tides in
terms of the tidal quality factor instead of a constant lag angle
$\phi$, or a constant time lag $\Delta t$. The specific dissipation
function is defined as follows \citep{Goldreich63}
\begin{equation}
Q\equiv\frac{2\pi
E^{*}}{\oint-\left(dE/dt\right)dt}=\frac{1}{\tan\phi}
\end{equation}
where $E^{*}$ in the numerator is the peak energy stored in tides 
during one tidal cycle, while the denominator represents the
energy dissipated over the cycle. When the phase angle $\phi$ (which
is twice the geometrical lag angle) is small, this is simplified as
$Q\sim1/\phi$, which implies a large $Q$ value. The estimated tidal
quality factors are $\gtrsim10^4$ for Jupiter and Neptune
\citep[e.g.,][]{Lainey09,Zhang08}, and even larger values are
expected for synchronized close-in exoplanets \citep{Ogilvie04}. On
the other hand, the usually adopted values are $\gtrsim10^6$ for
main sequence stars \citep[e.g.][]{Trilling98}. Thus, we are
generally interested in the case of $Q\gg1$, and the above
approximation is reasonable. Using the weak friction approximation
($\phi\sim\Delta t\sigma$), we can redefine $Q$ as
\begin{equation}
Q\equiv\frac{1}{\sigma\Delta t} \ .
\end{equation}

In the following sections, we study some limiting cases of tidal
frequencies. Before the spin-orbit synchronization, the tidal
dissipation is generally dominated by the semi-diurnal tide with the
forcing frequency of $\sigma=\left|2\omega-2n\right|$
\citep{FerrazMello08}. In this case, the tidal quality factor can be
written as
\begin{equation}
Q\sim\frac{1}{\Delta t \left|2\omega-2n\right|} \ .
\end{equation}
As the system approaches synchronization, the effect due to the
semi-diurnal tide diminishes, and the annual tide with
$\sigma=\left|2\omega-n\right|$ prevails \citep{FerrazMello08}. The
corresponding tidal quality factor can be written in a similar
manner to the above as
\begin{equation}
Q\sim\frac{1}{\Delta t \left|2\omega-n\right|} \ .
\end{equation}
Note that when $\omega=n$, the most efficient energy dissipation
occurs on the same timescale as the orbital period.

In Section~4 and 5, we rewrite eq.~\ref{eq1}-\ref{eq5} by assuming
that both stellar and planetary tidal quality factors change as
$Q\propto1/n$, while in Section~6, we adopt
$Q_*\propto1/\left|2\omega-2n\right|$ and
$Q_p\propto1/\left|2\omega-n\right|$. The former is chosen to
compare our results with the study of LWC09. For the latter, we are
implicitly assuming that close-in exoplanets are
pseudo-synchronized, while their host stars are not. We will see
that this is a reasonable approximation in Section~4.

Throughout this paper, we discuss our results in terms of the
modified tidal quality factor $Q'\equiv1.5Q/k_{2}$,
where $k_{2}$ is the Love number for the second-order harmonic
potential \citep{Love44}.

\section{Future Tidal Evolution of Transiting Planets}
\subsection{Two Evolutionary Paths for Darwin-Unstable Extrasolar Planets}
In this subsection, we revisit the tidal stability problem for
transiting extrasolar planets, and we show that there are two
distinctive evolutionary paths for Darwin-unstable systems.
Throughout this subsection, we assume that the total angular
momentum is conserved (i.e., magnetic braking is neglected). We take
into account the magnetic braking effects later in Section~4.3.

The existence and stability of tidal equilibrium states were
investigated by many authors
\citep[e.g.][]{Hut80,Peale86,Chandrasekhar87,Lai94} for binary
systems and the Solar System. Minimizing the total energy under the
constraint of conservation of total angular momentum, \cite{Hut80}
found that all equilibrium states are characterized by orbital
circularity, spin-orbit alignment, and synchronization of the
stellar rotation with the orbit. He also showed that equilibrium
states exist only when the total angular momentum of the system
$L_{\rm tot}$ is larger than some critical value $L_{\rm crit}$, and
that such equilibrium states are unstable when the orbital angular
momentum is less than 3 times the total spin angular momentum
($L_{\rm spin}/L_{\rm orb}>1/3$). In this paper, we are interested
in the ultimate fate of close-in exoplanets, and thus we call a
system "Darwin-stable", only when it is expected to evolve
ultimately toward a stable equilibrium state. All other
systems are called "Darwin-unstable".

\begin{figure}
\plotone{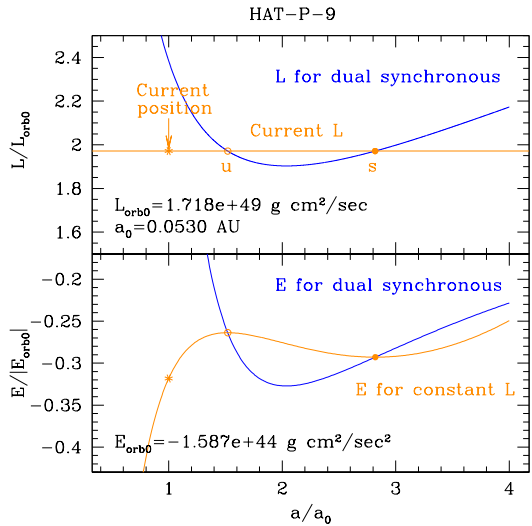} \caption{Tidal equilibrium curves as
a function of orbital separation for HAT-P-9. The total angular
momentum and total energy for dual synchronous states are plotted in
blue curves, while the corresponding curves with the constant
angular momentum are plotted in orange. The current values are
indicated as the star, while open and filled circles with u and s
denote unstable, and stable tidal equilibria, respectively. Although
HAT-P-9 has tidal equilibria, the system currently exists inside the
unstable tidal equilibrium state. Therefore, HAT-P-9 is Darwin
unstable and migrates toward the central star as the energy
dissipates.
 \label{figunst}}
\end{figure}

First, we check whether the stable tidal equilibrium states exist
for the currently known transiting planets listed in
Table~\ref{tab1}.
Tidal equilibrium can only exist when both primary and secondary,
with zero obliquities, are synchronously rotating with their orbital
motion. Under these constraints, the total angular momentum
\begin{displaymath}
L_{\rm tot}=L_{\rm orb}+C_*\omega_*+C_p\omega_p =L_{\rm orb}+(C_*+C_p)n
\end{displaymath}
has a minimum as a function of the semimajor axis $a$ when $L_{\rm orb}=3L_{\rm spin}$ and
\begin{displaymath}
L_{\rm tot}=L_{\rm crit}=4\left[\frac{G^2}{27}\frac{M_*^3M_p^3}{M_*+M_p}(C_*
+C_p)\right]^{\frac{1}{4}}.
\end{displaymath}
Here, $C=\alpha MR^{2}$ is the moment of inertia, and $L_{\rm
orb}=\frac{M_*M_p}{\sqrt{M_*+M_p}}\sqrt{Ga(1-e^2)}$ is the orbital
angular momentum. For $L_{\rm tot}<L_{\rm crit}$ there can be no
tidal equilibrium; for $L_{\rm tot}>L_{\rm crit}$ two equilibrium
states exist. The inner state ($L_{\rm orb}<3L_{\rm spin}$) is
unstable, so the only stable tidal equilibrium is the outer state,
which requires $L_{orb}>3L_{spin}$ \citep[e.g.,][also see the top
panel of Fig.~\ref{figunst}]{Hut80,Peale86}. A local example of dual
synchronous rotation is the Pluto-Charon system
\citep[e.g.,][]{Peale86}.
%
%
%

The results are summarized in Table~\ref{tab2}. As already pointed
out by LWC09, most systems have no tidal equilibrium states
(i.e.,~$L_{\rm tot}/L_{\rm crit}<1$), and thus are Darwin unstable
(i.e.,~the planet will eventually fall all the way to the Roche
limit of the central star, even when the total angular momentum is
strictly conserved). Note that, using the refined parameters in
\citet{Pal10}, even HAT-P-2, which was the only system with tidal
equilibria in LWC09, in fact now appears to have no such states
($L_{\rm tot}/L_{\rm crit}=0.995$).

There are several systems which have tidal equilibria  ($L_{\rm
tot}/L_{\rm crit}\gtrsim1$), and thus could be Darwin-stable,
namely, CoRoT-3, CoRoT-6, HAT-P-9, HD~80606, Kepler-8, WASP-7, and
WASP-17.
%
Among these, CoRoT-3, CoRoT-6, and HD~80606 are clearly evolving
toward the stable equilibrium state since they all have $L_{\rm
*,spin}/L_{\rm orb}<1/3$. Therefore, these systems are
Darwin-stable. Out of the systems with $L_{\rm *,spin}/L_{\rm
orb}>1/3$, WASP-17 is Darwin-unstable independent of the 1/3
criterion, since it is in a retrograde orbit ($L_{\rm orb}<0$) and
thus has $L_{\rm spin}>L_{\rm orb}$. For the other systems (HAT-P-9,
Kepler-8, and WASP-7), we checked their tidal equilibrium states as
in Fig.~\ref{figunst}. Here, we compare the total angular momentum
and total energy for dual synchronous state with the current values.
As clearly seen in the figure, HAT-P-9 is Darwin-unstable, since the
system is \emph{inside} the inner unstable equilibrium state. Most
of the angular momentum in the system is in the spin, and the planet
falls toward the central star as a result of tidal energy
dissipation. Similarly, Kepler-8 is Darwin-unstable since the system
is far inside the inner unstable equilibrium state. On the other
hand, WASP-7 exists just outside the inner unstable equilibrium
state, and thus is migrating outward, toward the stable tidal
equilibrium state. Similar plots for CoRoT-3, CoRoT-6, and HD~80606
reveal that CoRoT-3 and CoRoT-6 are migrating \emph{outward} toward
the stable equilibria, while HD~80606 is migrating inward toward the
stable state. For systems with non-zero eccentricities (CoRoT-3,
HD~80606, and WASP-17), we show the evolution explicitly in
Fig.~\ref{forwardQL09ae}. In short, we find that all planetary
systems in Table~\ref{tab1} except CoRoT-3, CoRoT-6, HD~80606, and
WASP-7 are Darwin-unstable.

\begin{figure}[t]
\plotone{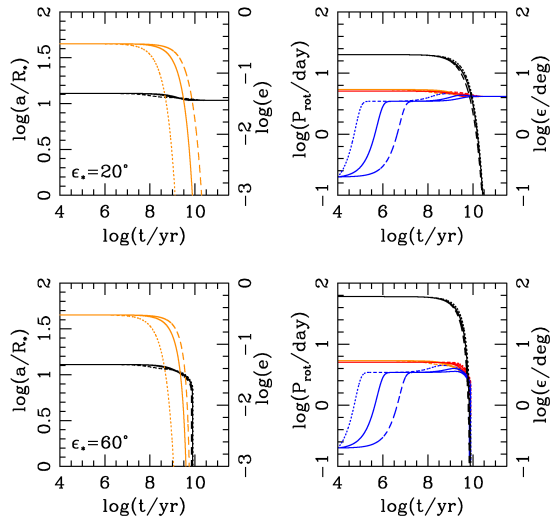} \caption{Tidal evolution of a planet with
$3M_J$ at $0.06\,$AU and $e=0.3$. In the left panels, black, and
orange curves show the evolution of semi-major axis and
eccentricity, respectively.  In the right panels, orange curves show
the evolution of orbital period, while blue and red curves show that
of planetary and stellar spin periods, respectively. The black
curves indicate the evolution of stellar obliquity. Dotted, solid,
and dashed curves correspond to cases with the same $Q'_{*,0}=10^6$
and different $Q'_{p,0}$ of $10^5$, $10^6$, and $10^7$,
respectively. Top panels show the cases with
$\epsilon_*=20^{\circ}$, which are Darwin-stable, and bottom panels
show the cases with $\epsilon_*=60^{\circ}$, which are
Darwin-unstable. \label{fig2}}
\end{figure}

The rest of the transiting systems have no tidal equilibrium
($L_{\rm tot}/L_{\rm crit}<1$) with the fiducial orbital and spin
parameters listed in Table~\ref{tab1}, and thus should be
Darwin-unstable. However, some of them are borderline systems with
$L_{\rm tot}/L_{\rm crit}\simeq 1$, and they could be either
Darwin-stable or -unstable within the observational uncertainties or
depending on the value of the unknown parameters (e.g., stellar
obliquity).
As an example, we show the tidal evolution of a hypothetical planet
with $M_p=3M_J$ and $R_p=1.2R_J$ orbiting a Sun-like star. We assume
the initial semimajor axis, eccentricity, and stellar velocity of
$a=0.06\,$AU, $e=0.3$, and $v_*=10\,$km/s. Such a system would have
a stable tidal equilibrium if the stellar obliquity is small,
because $L_{\rm tot}/L_{\rm crit}\sim1.040$ and $L_{\rm
*,spin}/L_{\rm orb}\sim0.193<1/3$.
In Fig.~\ref{fig2} we show the results of tidal evolution for two
different stellar obliquities of $\epsilon_*=20^{\circ}$ and
$60^{\circ}$. Here the tidal quality factors are scaled as
$Q'=Q'_{0}n_0/n$, where $0$ indicates the initial/current values.
This scaling ensures that our model is consistent with the constant
time lag model as shown in Section~3.1. For each obliquity, we
performed three different runs with the same $Q'_{*,0}=10^6$ and
different $Q'_{p,0}$ of $10^5$, $10^6$, and $10^7$. For the smaller
stellar obliquity ($\epsilon_*=20^{\circ}$), we find that the system
arrives at the stable tidal equilibrium as circularization,
synchronization, and alignment are achieved. On the other hand, for
the larger stellar obliquity ($\epsilon_*=60^{\circ}$), the system
turns out to be Darwin unstable, and the planet spirals into the
star on a $\sim10\,$Gyr timescale.
%
The examples of these borderline systems include HAT-P-2, WASP-10,
and XO-3. XO-3 was a Darwin-stable system with previously obtained
observed parameters, while HAT-P-2 and WASP-10 can be Darwin-stable
within the uncertainties as we see in Fig.~\ref{forwardQL09ae}.

The tidal evolution is not completely simple even for clearly
Darwin-unstable systems. The bottom panels of Fig.~\ref{fig2}
demonstrate that such systems can take either of two different
evolutionary paths, depending on the relative efficiency of tidal
dissipation inside the star and inside the planet. With the smaller
planetary tidal quality factors and thus with more efficient tidal
dissipation in the planet ($Q'_{p,0}=10^5$, and $10^6$ in the
figure, which correspond to dotted and solid curves, respectively),
the planetary orbit circularizes before the planet spirals into the
Roche limit ($\tau_{e}<\tau_{a}$), while with the larger planetary
tidal quality factors ($Q'_{p,0}=10^7$ in the figure, corresponding
to dashed curves), the circularization time becomes comparable to
the orbital decay time ($\tau_{e}\sim\tau_{a}$). The difference
occurs because the orbital decay time is largely determined by
dissipation inside the star, while the circularization time can be
determined by either dissipation inside the star, or that inside the
planet. This is apparent from the figure --- the orbital decay times
are similar for different $Q'_{p,0}$ values, which means that the
semi-major axis evolution is largely independent of tidal
dissipation in the planet, and instead is determined entirely by
dissipation in the star ($\tau_{a}\sim\tau_{a,*}$). On the other
hand, the eccentricity damping timescales change significantly for
different $Q'_{p,0}$ values, which suggests that dissipation in the
planet plays a significant role in circularization. However, the
circularization time is never longer than the orbital decay time.
Therefore, when the expected circularization time from $Q'_{p}$ is
longer than the orbital decay time, the eccentricity damping time is
also determined by the stellar tidal dissipation
($\tau_{e}\sim\tau_{e,*}$, which corresponds to $Q'_{p,0}=10^7$ case
in the figure).
We show that these results are generally true in Section~4.3.

Unfortunately, it is nontrivial to constrain tidal quality factors
and determine which type of evolution each system would follow. This
is because tidal quality factors depend sensitively on the detailed
interior structure of the body (either star or planet), as well as
the tidal forcing frequency and amplitude, and are unlikely to be
expressed as a simple constant value
\citep[e.g.,][]{Ogilvie04,Ogilvie07,Wu05b}. However, since we
observe a sharp eccentricity decline within $a\lesssim0.1\,$AU, and
many Darwin-unstable extrasolar planets are observed to be on nearly
circular orbits, it is likely that the eccentricity damping time
tends to be short compared to the orbital decay time
($\tau_{e}<\tau_{a}$). We discuss this issue further in Section~5.2.

In summary, we find that
there are two evolutionary paths for Darwin-unstable planets---the
"stellar-dissipation dominated" case in which all the parameters
except planetary spin evolve on a similar timescale
($\tau_{e}\sim\tau_{a}\sim\tau_{\epsilon_{*}}\sim\tau_{\omega_{*}}$),
and the "planetary-dissipation dominated" case in which
circularization occurs before any significant orbital decay
($\tau_{e}<\tau_{a}\sim\tau_{\epsilon_{*}}\sim\tau_{\omega_{*}}$).
In the former case, not only the evolution of stellar spin and
obliquity, but also that of semi-major axis and eccentricity are
controlled by the efficiency of tidal dissipation in the star. In
the latter case, the eccentricity damping is driven by dissipation
in the planet, while the orbital decay is largely determined by
dissipation in the star.

\subsection{Comparison with \cite{Levrard09}}
Note that, although the stellar-dissipation-driven case
($\tau_{e}\sim\tau_{a}$) looks similar to the one illustrated by
LWC09, there are fundamental differences. In Fig.~2 of LWC09, the
exponential eccentricity damping approximation, in which they only
integrate the planetary dissipation term in the eccentricity
evolution, shows a \textit{faster} damping time than the tidal
evolution involving both stellar and planetary energy dissipation.
Generally, however, this approximation should provide a timescale
comparable to, or longer than, what is obtained with the full
integration. This is because most planet-hosting stars are spinning
slower than the orbital frequency (see Table~\ref{tab2}), and thus
energy dissipation in the star accelerates the eccentricity damping,
rather than slowing it down \citep[see also][]{DobbsDixon04}.
Therefore, for most systems, the exponential damping approximation
involving only the dissipation in the planet should provide an
\textit{upper} limit for eccentricity damping.

\begin{figure}[t]
\plotone{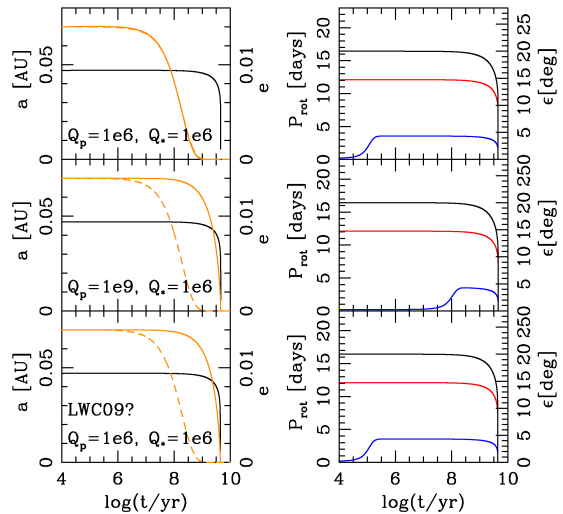}
\caption{Evolution of the orbital and spin parameters of HD~209458
(cf.~Fig.~2 of LWC09). Three different cases are shown from top to
bottom. In all cases, the integrations are stopped when the planet
formally hits the stellar surface. In the left panels, black and
orange curves represent the evolution of semi-major axis and
eccentricity, respectively. The dashed orange curve is the
exponential damping approximation. In the right panels, black, blue,
and red curves show the evolution of stellar obliquity, planetary,
and stellar spins, respectively. In top panels, we assume the same
initial conditions as LWC09, with $Q'_{p,0}=Q'_{*,0}=10^{6}$. The
evolution for all parameters but eccentricity looks similar to
LWC09's results. The eccentricity evolution follows the exponential
damping approximation. In middle panels, we use the same initial
conditions as in the top panels, but assume a less efficient tidal
damping in the planet ($Q'_{p,0}=10^{9}$). In this case, the
eccentricity evolution resembles the one in LWC09, but the planetary
spin-orbit synchronization occurs on a longer timescale, as
expected. In bottom panels, we use the same initial conditions as in
the top panels, but artificially multiply the eccentricity evolution
contribution from the planet by some small factor. Here, we recover
the results of LWC09. \label{fig3}}

\end{figure}

We demonstrate this in Fig.~\ref{fig3} by taking HD~209458 (the same
system considered in LWC09) as an example. Note however that, with
current data, HD~209458 has an orbital eccentricity consistent with
zero, and thus is not included in the analysis presented in the rest
of this paper. We adopt the same initial conditions and assumptions,
and use the same set of tidal equations as LWC09. Here the tidal
quality factors are initially $Q'_{p,0}=Q'_{*,0}=10^{6}$, and scale
as $Q'=Q'_{0}n_0/n$. Our result is shown in the top panel of
Fig.~\ref{fig3}. We find that the tidal evolution in Fig.~2 of LWC09
is recovered {\em except for the eccentricity}. In our results, the
eccentricity evolves on a timescale consistent with the exponential
eccentricity damping approximation (dashed line, which is completely
overlapped with solid line in the top left panel). In other words,
the eccentricity damps on the timescale determined by tidal
dissipation in the planet, as previously shown by many authors
\citep[e.g.,][]{Rasio96,DobbsDixon04,Mardling04}. For the middle
panels, we assume a less efficient tidal dissipation for the planet
($Q'_{p,0}=10^{9}$). In this case, we obtain an eccentricity
evolution similar to LWC09, but the planetary spin-orbit
synchronization, as expected, occurs more slowly. Note that the
dashed curve is the eccentricity damping approximation with
$Q'_{p,0}=10^{6}$. Finally, for the bottom panels, we reproduce
Fig.~2 of LWC09 by artificially reducing the planetary contribution
term in the eccentricity evolution equation (i.e.,~second term in
Eq.~\ref{eq2}), by a factor of $10^{3}$, i.e., completely
suppressing the eccentricity damping effect due to the planet
\footnote{Note that the choice of the factor is rather arbitrary.
The exact number can be anything as long as the planetary
contribution becomes negligible compared to the stellar
contribution.}.
As already mentioned, this discrepancy occurs because the numerical
results in LWC09 indeed underestimated the eccentricity damping in
the planet, since their code incorrectly had an additional factor of
n multiplied in the second term of Eq.~\ref{eq2} (B.\ Levrard,
private communication).

\subsection{Lifetimes of Transiting Planets on Eccentric Orbits}
In Section~4.1, we identify two characteristic evolutionary paths
for Darwin-unstable planets. We now study the tidal evolution of
eccentric transiting planets by integrating the tidal equations
forward in time with various tidal quality factors, and further
investigate the implications of these two paths.

We use the currently observed parameters as initial conditions (see
Table~\ref{tab1}). The orbital and stellar spin parameters are taken
from \emph{The Extrasolar Planets Encyclopedia}
(http://exoplanet.eu/), and references therein. For the planetary
spin period we assume a small initial value (0.2 days), but the
overall results are not affected by the exact choice of this value.
This is because the planetary spin carries a very small angular
momentum, and thus pseudo-synchronization with the orbit is achieved
very quickly (see also LWC09). For the stellar obliquity, we use the
observed projected value $\lambda$ when RM measurements are
available. Here, we implicitly assume that both the angle between
the stellar spin axis and the line of sight ($i_*$) and the angle
between the orbital angular momentum and the line of sight ($i_o$)
are $\simeq 90^{\circ}$. Thus, from $\cos\epsilon_*=\sin
i_*\cos\lambda\sin i_o+\cos i_*\cos i_o$ \citep{Fabrycky09}, the
stellar obliquity becomes comparable to the projected one
($\epsilon_*\simeq \lambda$). For systems without RM measurements,
we assume an initial stellar obliquity $\epsilon_{*,0}=2^{\circ}$.
This choice is rather arbitrary, but is motivated by the typical
projected obliquity observed (see Table~\ref{tab2}).

\begin{figure}[t]
\epsscale{0.9}\plotone{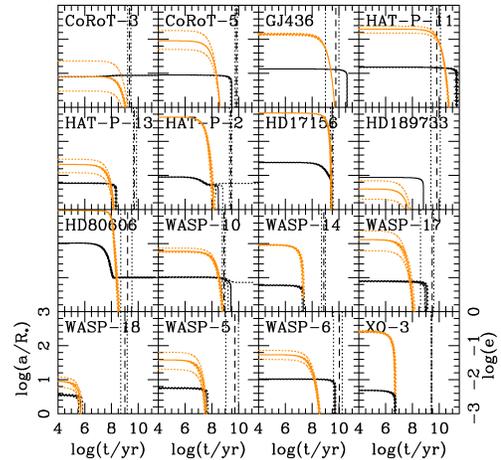} \caption{The
evolution of semimajor axis (black curves) and eccentricity (orange
curves) for $Q'_{*,0}=Q'_{p,0}=10^6$. Tidal quality factors scale as
$Q'=Q'_0n_0/n$, and magnetic braking is not included. Solid curves
correspond to the fiducial values, while dotted curves correspond to
four different combinations of maximum and minimum semimajor axis
and eccentricity, allowed within the uncertainties. Vertical lines
show the age of each system (dashed lines) with uncertainties
(dotted lines). As expected from Section~4.1, CoRoT-3 and HD~80606
arrive at their stable tidal equilibrium, while the other planets
spiral into the central star. This figure clearly demonstrates that
different tidal quality factors apply to different systems, since
some planets fall within the Roche limit of their stars on time
scales much shorter than the age of the star with a common value of
$Q'_{*,0}$. \label{forwardQL09ae}}
\end{figure}

First, we show the tidal evolution for typical tidal quality factors
($Q'_{*,0}=Q'_{p,0}=10^6$), with the scaling of $Q'=Q'_0n_0/n$, and
without magnetic braking effect. Our goal here is to show how
different the evolution can be within observational uncertainties.
We exclude WASP-12 and HAT-P-1 from our analysis because of the
uncertainty in their age.
The evolution of semimajor axis and eccentricity of each system is
shown in Fig.~\ref{forwardQL09ae}. In each panel, we show the
results of 5 different integrations. Solid curves represent the
results with fiducial values of $a$ and $e$, while dotted curves
correspond to 4 different combinations of maximum and minimum $a$
and $e$ values within their error bars. As expected from Section~4.1, 
the Darwin-stable systems CoRoT-3 and HD~80606 migrate outward
and inward, respectively, and arrive at their tidal equilibrium,
with orbital decay eventually stopping. The borderline systems
HAT-P-2 and WASP-10 are likely Darwin-unstable, but may arrive at a
stable tidal equilibrium within the observed uncertainties. Thus, it
is important to know the orbital and spin parameters as well as
possible to determine the final fate of these borderline systems.
The other systems are definitely Darwin-unstable within the current
observational accuracy, and their planets spiral toward the central
star on different timescales. Vertical lines indicate the estimated
ages with uncertainties. With $Q'_{*,0}=Q'_{p,0}=10^6$ some systems
undergo orbital decay too quickly to be compatible with their likely
age (e.g., WASP-18, XO-3). Therefore, these results clearly imply
that a single set of values for the tidal quality factors cannot
reasonably apply to all systems \citep[also see, for
example,][]{Matsumura08}.

\begin{figure}
\plotone{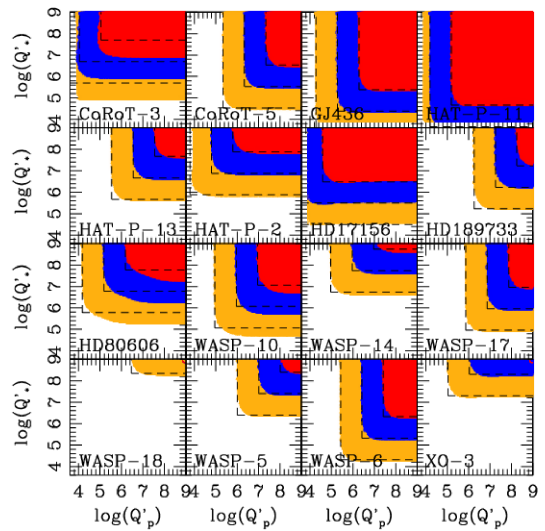}

\caption{Combinations of stellar and planetary tidal quality factors
which keep a non-zero eccentricity and allow survival of the planet
in forward integration of the tidal equations for $0.1$, $1$, and
$10\,$Gyr (orange, blue, and red regions, respectively). Magnetic
braking is not included, and tidal quality factors change as
$Q'=Q'_0n_0/n$. Vertical, and horizontal dashed lines are determined
by assuming $(de/dt)\sim(de/dt)_p$, and $(da/dt)\sim(da/dt)_*$,
respectively (see text for details). The system's lifetime is
largely determined by the tidal dissipation in the star, and the
circularization by that in the planet. \label{fig4}}

\end{figure}

We repeated the above calculations for various initial stellar and
planetary tidal quality factors ranging over the interval $10^4\leq
Q'_{0}\leq 10^9$, with the scaling of $Q'=Q'_0n_0/n$.
%
In Fig.~\ref{fig4}, we show the range of values for the tidal
quality factors that allow planets to survive ($a\gtrsim R_*$), and
stay eccentric ($e\gtrsim0.001$) for $0.1$, $1$, and $10\,$Gyr. Here
magnetic braking is not included. For Darwin-stable systems
(CoRoT-3, and HD~80606 in our list), the minimum stellar and
planetary tidal quality factors correspond to the orbital
circularization times. For all Darwin-unstable transiting systems
with noncircular orbits, we find the same trend as in Section~4.1:
the circularization time is largely determined by the dissipation in
the planet, while the survival time is largely determined by
dissipation inside the star. In other words, for Darwin-unstable
planets, the minimum planetary tidal quality factors can be inferred
from the circularization time, while the minimum stellar tidal
quality factors can be inferred from the orbital decay time. We
demonstrate this below.

The approximate minimum planetary tidal quality factor that allows a
planet to keep a non-circular orbit ($e\gtrsim0.001$) for a certain
time (in our examples, $0.1-10\,$Gyr) can be determined by assuming
that the eccentricity evolution depends only on tidal dissipation
inside the planet $(de/dt)\simeq (de/dt)_p$. We rewrite
Eq.~\ref{eq2} as follows by assuming pseudo-synchronization of the
planetary spin ($d\omega_p/dt=0$), as well as conservation of
angular momentum ($a(1-e^2)=\,$const):
\begin{equation}
\frac{de}{dt} \sim \frac{81}{2}\frac{n_0}{Q'_{p,0}}\frac{M_{*}}{M_{p}}\frac{R_{p}^{5}}{a_0^{5}}(1-e_0^2)^{-8}
e(1-e^{2})^{3/2}\left[\frac{11}{18}\frac{f_{2}(e^{2})f_{4}(e^{2})}{f_{5}(e^{2})}-f_{3}(e^{2})\right].  \label{dedtapprox}
\end{equation}
By integrating the above equation from the currently observed
eccentricity to $e=0.001$, and solving for $Q'_{p,0}$, we obtain the
minimum planetary tidal quality factors to keep a non-circular orbit
for $0.1$, $1$, and $10\,$Gyr, respectively. These values are
plotted as the vertical dashed lines in Fig.~\ref{fig4}.

Similarly, we can determine the approximate minimum stellar tidal
quality factor that allows a planet to survive for a certain time
before plunging into the central star, by assuming that the
semimajor axis evolution depends only on tidal dissipation inside
the star $(da/dt)\simeq (da/dt)_*$.  Note that this is a reasonable
approximation when the pseudo-synchronization of the planetary spin
is achieved, and the orbit is nearly circular. We rewrite
Eq.~\ref{eq1} as follows by setting $e=0$,
\begin{equation}
\frac{da}{dt} \sim \frac{9}{Q'_{*,0}}\frac{M_{p}}{M_{*}}\frac{R_{*}^{5}}{a^{4}}\left(\frac{a_0}{a}\right)^{3/2} \left[\omega_{*,0}\cos\epsilon_{*,0}-n\right]. \label{dadtapprox}
\end{equation}
We integrate the above equation from $a=a_0(1-e_0^2)$ to $R_*$, and
solve for $Q'_{*,0}$ to obtain the horizontal dashed lines. Here, we
make use of the fact that the difference in eccentricity damping
times does not strongly affect the orbital decay time, and assume
that the orbital decay time of any eccentric Darwin-unstable system
can be well described by that of a system with equal angular
momentum and a circular orbit.

\begin{figure}
\plottwo{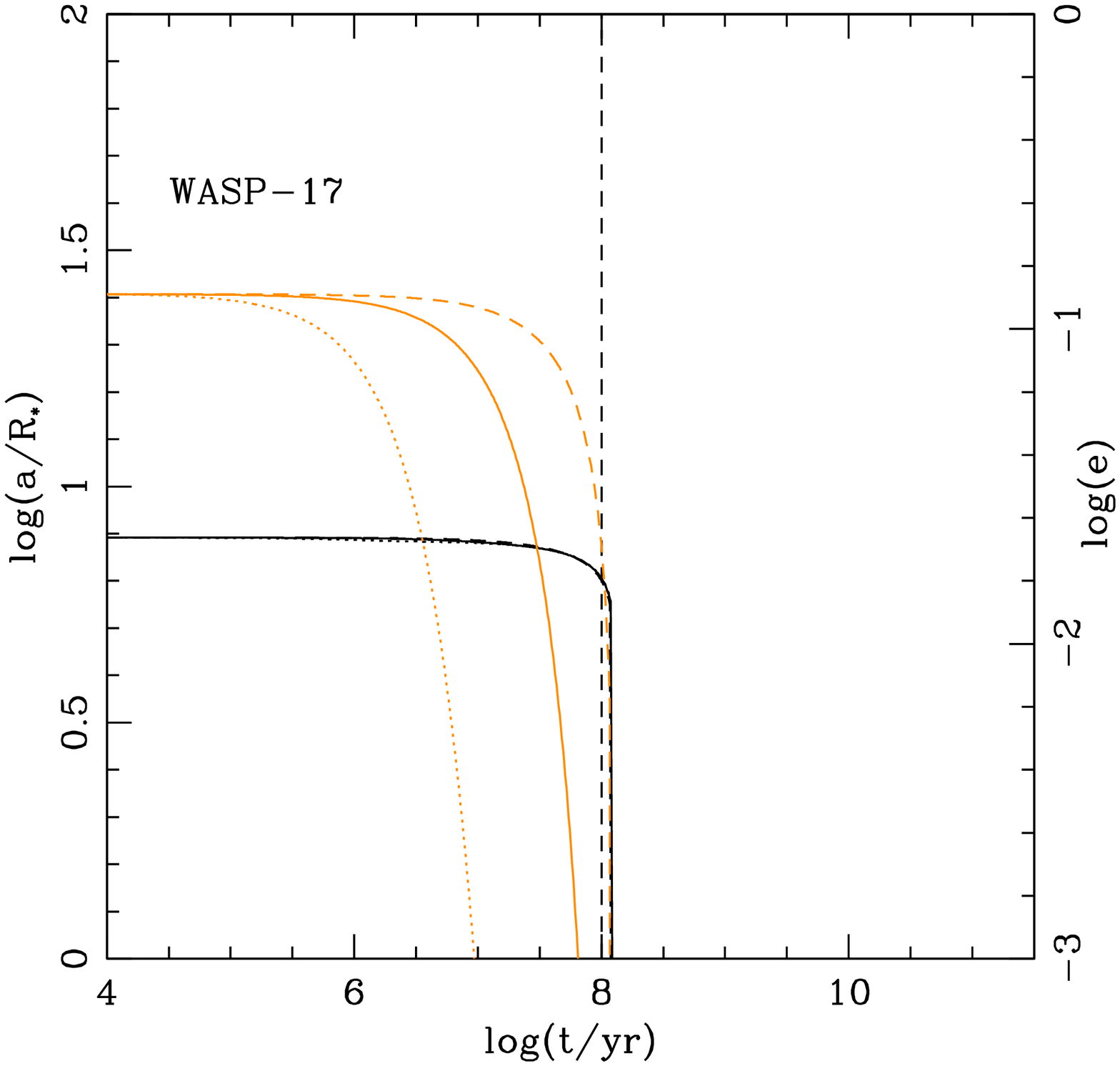}{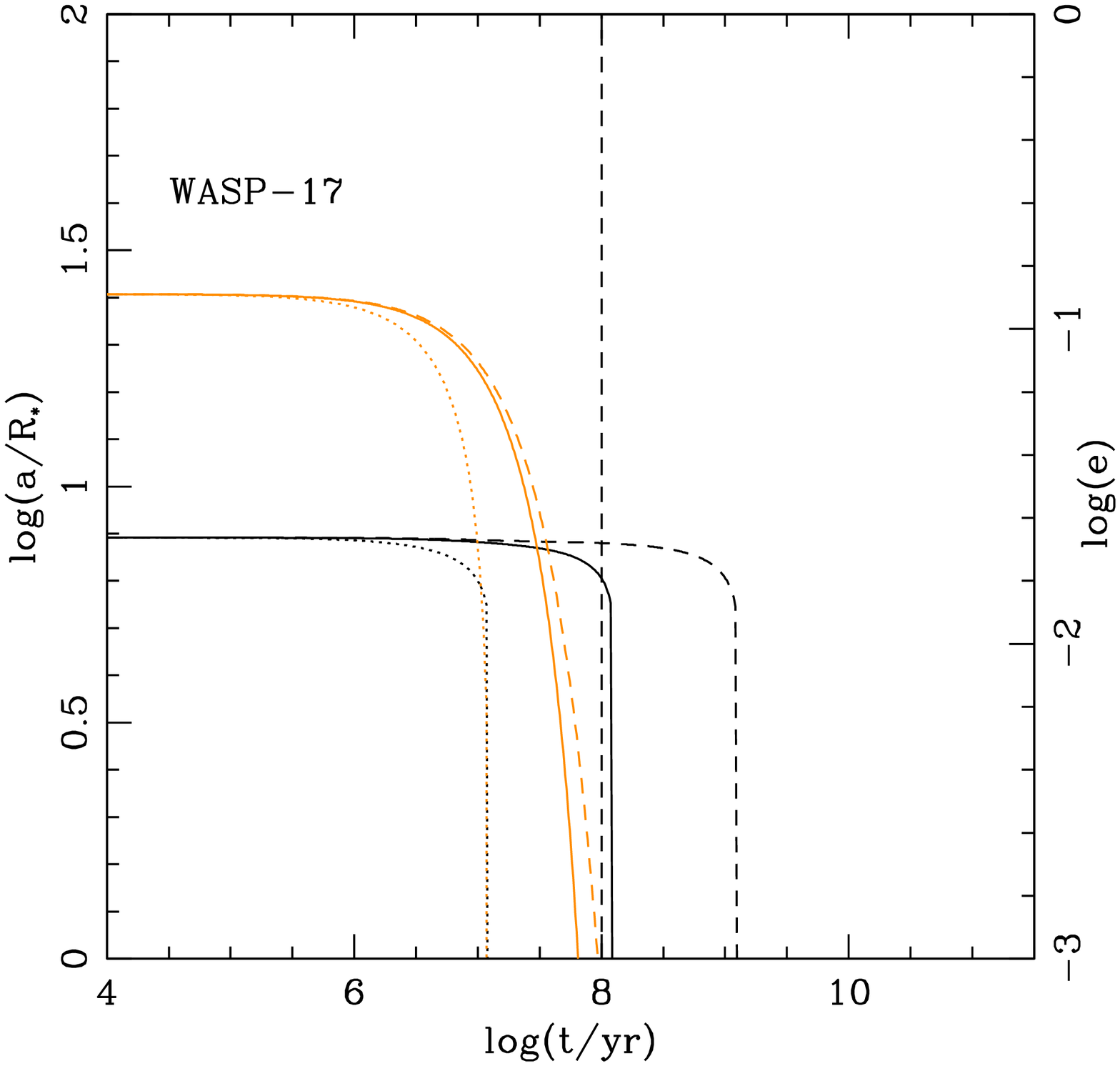}
\caption{Tidal evolution of WASP-17 with different initial tidal
quality factors along the dashed lines in Fig.~\ref{fig4}. Black and
orange curves correspond to semimajor axis and eccentricity
evolutions, respectively. The vertical dashed lines are drawn at
$0.1\,$Gyr for comparison. Left: Different initial conditions along
the lowermost horizontal dashed line in Fig.~\ref{fig4} that
indicates the survival time of $0.1\,$Gyr. Three runs with the same
initial stellar tidal quality factor $Q'_{*,0}=9.13\times10^4$, and
different initial planetary tidal quality factors are shown. Dotted,
solid, and dashed curves correspond to $Q'_{p,0}=7.43\times10^4$,
$7.43\times10^5$, and $7.43\times10^6$, respectively. Orbital decay
time is determined by the tidal dissipation in the star, since the
decay time does not change for different $Q'_{p,0}$ values. For
$Q'_{p,0}>7.43\times10^5$, it is clear that the eccentricity damps
on a similar timescale to the orbital decay.
Right: Different initial conditions along the leftmost vertical
dashed line in Fig.~\ref{fig4} that indicates the circularization
time of $0.1\,$Gyr. Three runs with the same initial planetary tidal
quality factor $Q'_{p,0}=7.43\times10^5$, and different initial
stellar tidal quality factors are shown. Dotted, solid, and dashed
curves correspond to $Q'_{*,0}=9.13\times10^3$, $9.13\times10^4$,
and $9.13\times10^5$, respectively. For $Q'_{*,0}<9.13\times10^4$,
both the orbital decay and the circularization times are much
shorter than $0.1\,$Gyr. For $Q'_{*,0}>9.13\times10^4$, both become
comparable to, or longer than $0.1\,$Gyr. \label{forwardQL09WASP17}
}
\end{figure}

As seen in Fig.~\ref{fig4}, the agreement of both horizontal and
vertical dashed lines with the calculations for Darwin-unstable
systems is very good. Since Darwin-stable systems (CoRoT-3 and
HD~80606) arrive at the stable tidal equilibria and stop migrating,
the horizontal dashed lines for these systems significantly differ
from the calculated results.
Now we present some example runs along horizontal and vertical lines
for WASP-17 to better understand their implications. The left panel
of Fig.~\ref{forwardQL09WASP17} shows three runs along the lowermost
horizontal line that corresponds to the survival time of $0.1\,$Gyr.
Dotted, solid, and dashed curves show the evolutions with the same
initial stellar tidal quality factor $Q'_{*,0}=9.13\times10^4$, and
different initial planetary tidal quality factors
$Q'_{p,0}=7.43\times10^4$, $7.43\times10^5$, and $7.43\times10^6$,
respectively. The figure confirms that the orbital decay time is
largely determined by the tidal dissipation in the star, since there
is no obvious difference depending on $Q'_{p,0}$ values. At the
vertex of vertical and horizontal dashed lines in Fig.~\ref{fig4}
(i.e.,~$(Q'_{*,0},\,Q'_{p,0})=(9.13\times10^4,\,7.43\times10^5)$),
we find that the semimajor axis and eccentricity damp roughly on the
similar timescale ($\tau_e\sim\tau_a\sim0.1\,$Gyr). For the smaller
$Q'_{p,0}$, the eccentricity damps much faster than the orbital
decay ($\tau_e<\tau_a\sim0.1\,$Gyr), while for the larger
$Q'_{p,0}$, the eccentricity damps slower than it is expected from
the other two curves, and on a similar timescale to the orbital decay
($\tau_e\sim\tau_a\sim0.1\,$Gyr).
Similarly, the right panel of Fig.~\ref{forwardQL09WASP17} shows
three runs with the same initial planetary tidal quality factor
$Q'_{p,0}=7.43\times10^5$, and different initial stellar tidal
quality factors $Q'_{*,0}=9.13\times10^3$, $9.13\times10^4$, and
$9.13\times10^5$. When $Q'_{*,0}$ is smaller than the vertex value,
the orbit decays much faster than $0.1\,$Gyr, and the
circularization happens on the similar timescale
($\tau_e\sim\tau_a<0.1\,$Gyr). On the other hand, when $Q'_{*,0}$ is
larger than the vertex value, the orbit decays much slower than
$0.1\,$Gyr, and the circularization time is shorter than the orbital
decay time ($\tau_e\sim0.1\,$Gyr$<\tau_a$). The figure also implies
that the orbital circularization time is largely determined by the
dissipation in the planet unless $\tau_e\sim\tau_a$. Thus, we find
that $\tau_e\sim\tau_a$ is a good approximation along the horizontal
dashed lines, while $\tau_e<\tau_a$ is a good approximation along
the vertical lines. In other words, the region below the diagonal
line drawn by connecting the vertices of vertical and horizontal
lines is the stellar-dissipation-dominated region, while the region
above the line is the planetary-dissipation-dominated region.

\begin{figure}
\plotone{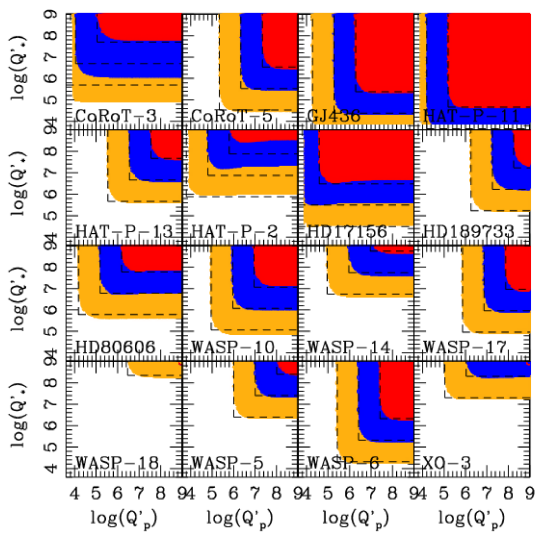}
\caption{Same as Fig.~\ref{fig4}, but with the effects of magnetic
braking included. There is very little difference for the future
tidal evolutions with and without magnetic braking. Vertical and
horizontal dashed lines are the same as in Fig.~\ref{fig4}.
\label{forwardmbQL09all}}

\end{figure}

We repeat these integrations by including magnetic braking effects.
As can be seen in Fig.~\ref{forwardmbQL09all}, there is little
difference between the cases with and without magnetic braking. Note
that dashed lines are the same as the ones in Fig.~\ref{fig4}, and
thus do not take account of the magnetic braking effects.
\cite{Barker09} suggested that the effect of magnetic braking can be
important for systems with rapidly spinning stars ($\omega_*
\cos\epsilon/n\gg1$). From Table~\ref{tab2}, there are five such
systems (CoRoT-3, CoRoT-6, HAT-P-2, HD~80606, and WASP-7). Among
them, CoRoT-3, CoRoT-6, HD~80606, and WASP-7 are Darwin-stable
systems, while HAT-P-2 is a borderline case that can be either
Darwin-stable or unstable within observational uncertainties.  Out
of these systems with fast spinning stars, CoRoT-3, HAT-P-2, and
HD~80606 have eccentric planets and are shown in
Fig.~\ref{forwardmbQL09all}. Excluding Darwin-stable cases (CoRoT-3
and HD~80606), HAT-P-2 indeed shows a significant difference between
Fig.~\ref{fig4} and \ref{forwardmbQL09all} for the $1$ and $10\,$Gyr
cases.

\section{Past Tidal Evolution of Transiting Planets}
\subsection{Observational Implications on the Origins of Close-in Planets}
In Section~1, we mentioned the two main scenarios to form close-in
planets --- planet migration in a disk, and tidal circularization of
an eccentric orbit, which may be obtained as a result of
planet--planet scattering or Kozai-type perturbations. It is
nontrivial to determine which formation mechanism dominates, but
there are at least a few observational indications that support the
second scenario.

One of them relates to the orbital distribution of planetary
systems. \citet{Wright09} compared the properties of multiple-planet
systems with single-planet systems (i.e.,~with no obvious additional
giant planet), and showed that their semi-major axis distributions
are different. While single-planet systems have a double-peaked
distribution, which is characterized by a pile-up of giant planets
between $0.03$ and $0.07\,$AU (the so-called ``3-day pileup'') as
well as a jump in the number of planets beyond $1\,$AU,
multiple-planet systems have a much more uniform distribution. They
also pointed out that the occurrence of close-in ($a<0.07\,$AU)
planets is lower for multiple-planet systems, and that planets
beyond $0.1\,$AU in multiple-planet systems exhibit somewhat smaller
eccentricities compared to the corresponding single ones. If
confirmed by future observations, this trend would favor
planet--planet interaction scenarios over a migration one, because
there is no obvious reason why the orbital distributions of single-
and multiple-planet systems should be different for planet
migration. On the other hand, gravitational interactions combined
with tidal circularization  may be able to explain such a
difference, because strong gravitational interactions tend to
disrupt the system, and thus currently observed multiple-planet
systems are unlikely to have been strongly perturbed by
stellar/planetary companions, and/or to have undergone catastrophic
scattering events.

Another indication comes from the similarity between the stellar
obliquity distribution derived from the observed systems, and the
distribution predicted by the Kozai migration scenario
\citep{Fabrycky07,Wu07}. \cite{Triaud10ap} observed the RM effect
for six transiting hot Jupiters, and found that four of their
targets appear to be in retrograde orbits with a projected stellar
obliquity $>90^{\circ}$. Combining the previous 20 systems with such
measurements, they pointed out that 8 out of 26 systems are clearly
misaligned, and that 5 out of 8 misaligned systems exhibit
retrograde orbits. They also derived the stellar obliquity
distribution by assuming an isotropic distribution of the stellar
spin with respect to the line of sight, and found that the
distribution closely matches that expected from the Kozai migration
scenario \citep{Fabrycky07,Wu07}. \cite{Fabrycky07} and \cite{Wu07}
independently studied the possibility of forming a close-in planet
by considering the combined effects of secular perturbations due to
a highly-inclined distant companion star (i.e.,~Kozai-type
perturbations) and tidal interactions with the central star. In this
scenario, a Jupiter-mass planet which was initially on a
nearly-circular orbit at $\sim5\,$AU can become a hot Jupiter. The
mechanism involves a companion star on a highly-inclined orbit
($\simeq 90^{\circ}$), which perturbs the planetary orbit and
increases its eccentricity. Once the pericenter distance of the
planet becomes small enough for tidal interactions with the central
star to be important, energy dissipation leads to circularization of
the planetary orbit, and eventually to formation of a hot Jupiter
with a small, or nearly zero eccentricity. They found that hot
Jupiters formed this way tend to be in misaligned orbits, and
frequently in retrograde orbits. Planet--planet interactions around
a single star (without a binary companion) could also form hot
Jupiters via Kozai migration \citep{Nagasawa08}.

Yet another clue regarding the origin of close-in planets is related
to the above scenario, and comes from the inner edge of the orbital
distribution for hot Jupiters. \citet{Ford06} proposed that the
observed inner cutoff for hot Jupiters is defined not by an orbital
period, but by a tidal limit, and studied such a cutoff of the
distribution of close-in giant planets in the mass-period diagram by
performing a Bayesian analysis. Assuming that a slope in such a
diagram follows the Roche limit, they found that the observations
suggest an inner cutoff close to {\em twice} the Roche limit. This
can be explained naturally if the planetary orbits were initially
highly eccentric, and later circularized via tidal dissipation while
conserving orbital angular momentum. They suggested that this result
is inconsistent with the migration scenario, because migration would
lead to an inner edge right at the Roche limit (a factor of 2
further in than what is observed).

\begin{figure}
\plotone{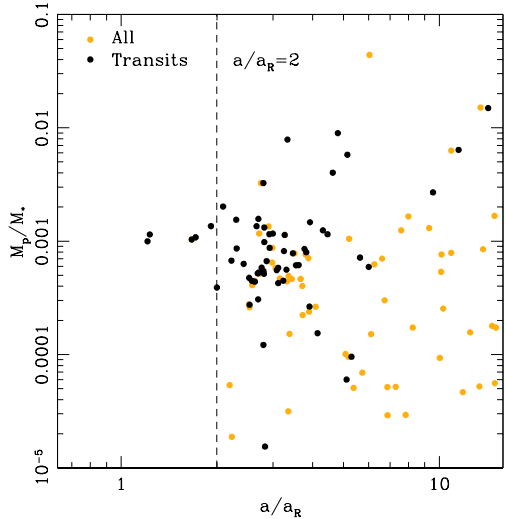}
%
\caption{Planetary and stellar mass ratio as a function of
semi-major axis normalized to the Roche-limit distance. For
non-transiting planets (orange circles) without planetary radius
information, we assume a Jupiter radius for planets with mass larger
than $0.1\, M_{J}$, and a Neptune radius for planets with smaller
mass.
Most close-in planets lie beyond 2 times their Roche Limit. One
planet, WASP-12b, has a non-zero eccentricity, and a very small
semi-major axis ($a\simeq1.3a_{R}$). \label{fig1}}

\end{figure}

In Fig.~\ref{fig1}, we extend the work of \cite{Ford06} by including
all more recently discovered planets, and we plot planetary and
stellar mass ratio against semi-major axis in terms of the Roche
limit separation \citep{Paczynski71}
\begin{equation}
a_{\rm R}=\frac{3}{2}R_p\left(\frac{M_p}{3\left(M_*+M_p\right)}\right)^{-1/3}
\sim \frac{R_{p}}{0.462}\left(\frac{M_{p}}{M_{*}}\right)^{-1/3} \ .
\end{equation}
Here, we assume that the Roche radius of the planet, which is
defined so that its spherical volume is equal to the volume within
the Roche lobe, is equal to the planetary radius. Thus, the Roche
limit separation used here is the \emph{lower} limit. For
non-transiting planets without a measured planetary radius (orange
circles), we assume a Jupiter radius for planets with mass larger
than $0.1\, M_{J}$, and a Neptune radius for less massive planets.
It is obvious that most planets still exist beyond twice the Roche
limit (vertical dashed line). However, there are now 5 planets which
lie within this limit (OGLE-TR-56, CoRoT-1, WASP-4, WASP-19, and
WASP-12) with $a/a_{\rm R}\sim1.70$, 1.67, 1.66, 1.30, and 1.24,
respectively. We need to assess whether these planets will change
the claim by \cite{Ford06}, and support the migration scenario over
the scattering/Kozai-cycle scenario. Note that the two recently
discovered ``extreme'' close-in planets, with orbital periods less
than 1 day, CoRoT-7~b and WASP-18~b, have $a/a_{\rm R}\sim2.76$, and
3.52, respectively.

The existence of at least some of these planets inside $2a_R$ may
still be explained as a result of tidal circularization of an
eccentric orbit. One of the possibilities is that the orbits of
these planets were originally circularized \emph{beyond} twice the
Roche limit, but the planets have migrated inward after
circularization due to tidal dissipation in the star. All of the
planets within $2a_R$ can be potentially explained this way.
Another possibility is that their orbits used to have a pericenter
distance close to the Roche limit, but the initial eccentricity was
smaller, $e\simeq0.7$. In such a case, the expected period of the
circularized orbit would be comparable to the current orbital period
for OGLE-TR-56, CoRoT-1, and WASP-4, since they all have $a/a_{\rm
R}\sim1.7$. However, systems with smaller ratios of semimajor axis
to Roche limit separation (WASP-12 and WASP-19) are unlikely to be
explained this way. This is because their current semimajor axes
($a/a_{\rm R}\simeq 1.24$ and $1.30$) would demand small initial
eccentricities ($e_i\sim0.24$ and $0.30$), and thus small initial
semi-major axes ($a_0\simeq 0.024\,$AU and $0.018\,$AU,
respectively). This means that they have to be either born at such
close-in locations, or scattered into such an orbit, which would be
very difficult.
%
Yet another possibility is that these planets used to have a smaller
Roche limit separation, due to a larger mass, or a smaller radius.
The orbits of these planets might have been circularized as the
planet lost mass \citep[e.g.][]{Lammer03,Lecavelier07}, or the
planetary radius expanded due to tidal heating
\citep[e.g.][]{Bodenheimer01,Gu03}. For OGLE-TR-56, CoRoT-1, WASP-4,
WASP-19, and WASP-12 to be circularized at twice the Roche limit,
either the past planetary masses must have been $2.13$, $1.78$,
$1.53$, $4.19$, and $5.98\,M_J$, respectively, or the past planetary
radii must have been $1.18$, $1.24$, $1.21$, $0.829$, and
$1.11\,R_J$, respectively. Since the mass-loss rate can only be at
most $\sim10\%$, even for a low-density planet like WASP-12
\citep{Lammer09}, it is unlikely that a larger mass in the past
could be the correct explanation. On the other hand, tidal inflation
of the planetary radius is a transient phenomenon
\citep[e.g.][]{Ibgui09}. To explain the current orbital radius by
invoking a smaller planetary radius in the past, we have to catch
the planet just as its radius is inflating.  Such a scenario may be
possible, but appears unlikely. Interestingly, CoRoT-1 is observed
to be strongly misaligned with $\lambda=-77\pm11\,$degrees
\citep{Pont10}, which suggests a violent origin, while WASP-4 has a
stellar obliquity of $4^{+34}_{-43}\,$degrees \citep{Triaud10ap},
which is consistent with alignment. We urge observers to determine
the projected stellar obliquity for OGLE-TR-56, WASP-19, and
WASP-12. Alignment does not necessarily support the planet migration
scenario over the violent origin, but misalignment would clearly
imply a significant past dynamical interaction involving scattering
or Kozai cycles.
%

\subsection{Past Evolution of Transiting Planets}
As we pointed out in the previous subsection, planet--planet (or
planet--companion star) interactions may well be responsible for the
majority of close-in planets. Now we further explore this
possibility by performing integrations of the tidal evolution
equations {\em backward} in time. Particular focus is on the
differences in evolution of Darwin-unstable planets depending on two
evolution paths.

\cite{Jackson08} first performed such a study, and estimated the
``initial'' eccentricity distribution of close-in planets. They
found that this distribution agrees well with the observed
eccentricity distribution of exoplanets on wider orbits, and they
proposed that most currently observed close-in planets had
considerably wider, and more eccentric orbits in the past. However,
their study had many uncertainties. First, the possible effects of
other planets and binary companions were neglected.  As we pointed
out in Section~2, these effects are most likely unimportant at
present, but they could well have been playing a crucial role in the
past, especially if another object was responsible for
Kozai-type perturbations, or gravitational scattering.
Second, the evolution of stars and planets is neglected.   For
example, both planetary and stellar radii may have been
significantly different in the past. Moreover, many stars could have
lost a large amount of their spin angular momentum via magnetic
braking. Finally, such backward dynamical integrations of evolution
equations with energy dissipation are known to be diverging, and
thus small changes in initial conditions can lead to much larger
changes in the calculated ``initial'' values.

\begin{figure}
\plotone{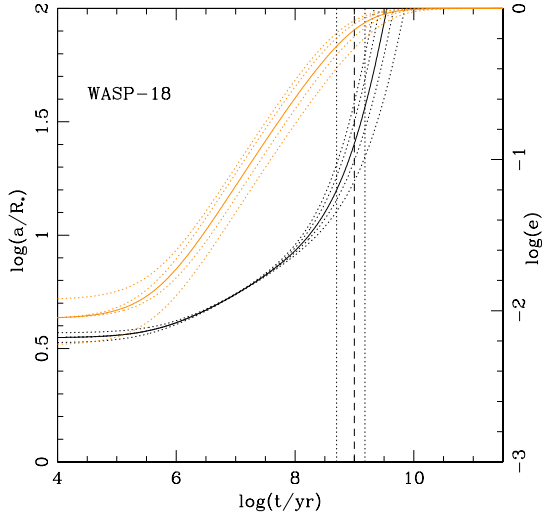} \caption{The backward
evolution of semi-major axis $a$ (black curves) and eccentricity $e$
(orange curves) for WASP-18 with $Q'_{*,0}=Q'_{p,0}=10^6$. Solid
curves are the nominal values, and dotted curves show four
independent runs with different combinations of $a$ and $e$ within
uncertainties. Vertical lines show the estimated stellar age with
uncertainties. \label{backwardQL09WASP18}}
\end{figure}

\begin{figure}
\plotone{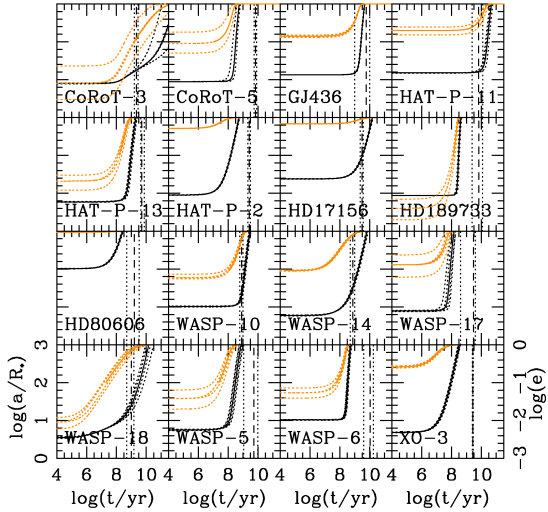} \caption{Same as the
previous figure for other transiting systems.  The evolution of
semi-major axis and eccentricity is shown with
$Q'_{*,0}=Q'_{p,0}=10^6$. \label{backwardQL09ae}}
\end{figure}

With these caveats in mind, we now re-examine the possible past
histories of transiting planets. Since Eq.~\ref{eq1}-\ref{eq5} are
time-invariant, we can study the past evolution by integrating the
differential equations ``backward in time'' (i.e.,~by taking the
negative of these differential equations and integrating them from
$t=0$ to $\tau_{\rm age}$). For simplicity, we assume that the
planetary spin is pseudo-synchronized with the orbit at all times
(i.e.,~$d\omega_{p}/dt=0$). First, we show typical results for
WASP-18 in Fig.~\ref{backwardQL09WASP18}. Here, the initial tidal
quality factors are $Q'_{*,0}=Q'_{p,0}=10^6$ with the scaling of
$Q'=Q'_0n_0/n$, and the magnetic braking effect is neglected. As in
Fig.~\ref{forwardQL09ae}, solid curves show the results of backward
evolution with the fiducial orbital parameters, while four dotted
curves correspond to different combinations of maximum and minimum
semimajor axis and eccentricity within uncertainties. Vertical
dashed and dotted lines correspond to the estimated stellar age with
uncertainties. As expected, both semimajor axis and eccentricity
values differ significantly within uncertainties at the zero age
(vertical dashed line), or at any specific time within the age
uncertainties (between vertical dotted lines). However, we find that
the fiducial case still provides a representative evolutionary path
within the stellar age uncertainties.
Fig.~\ref{backwardQL09ae} presents similar backward evolutions for
all the systems shown in Fig.~\ref{forwardQL09ae}. It is clear that,
except the Darwin-stable CoRoT-3, WASP-18 has the largest spread in
orbital parameters within uncertainties. The backward evolution of
the other systems is largely independent of the exact values of
orbital parameters.

Now we repeat backward integrations of the tidal equations by
adopting various combinations of initial stellar and planetary tidal
quality factors ranging over $10^4\leq Q'_{0}\leq 10^9$. This allows
us to study the ``initial'' orbital properties at the zero stellar
age ($\tau_{age}=0$) within the estimated age uncertainties. As in
Section~4.3, we also assumed that each system initially has the
currently observed orbital and rotational parameters.

\begin{figure}
\plotone{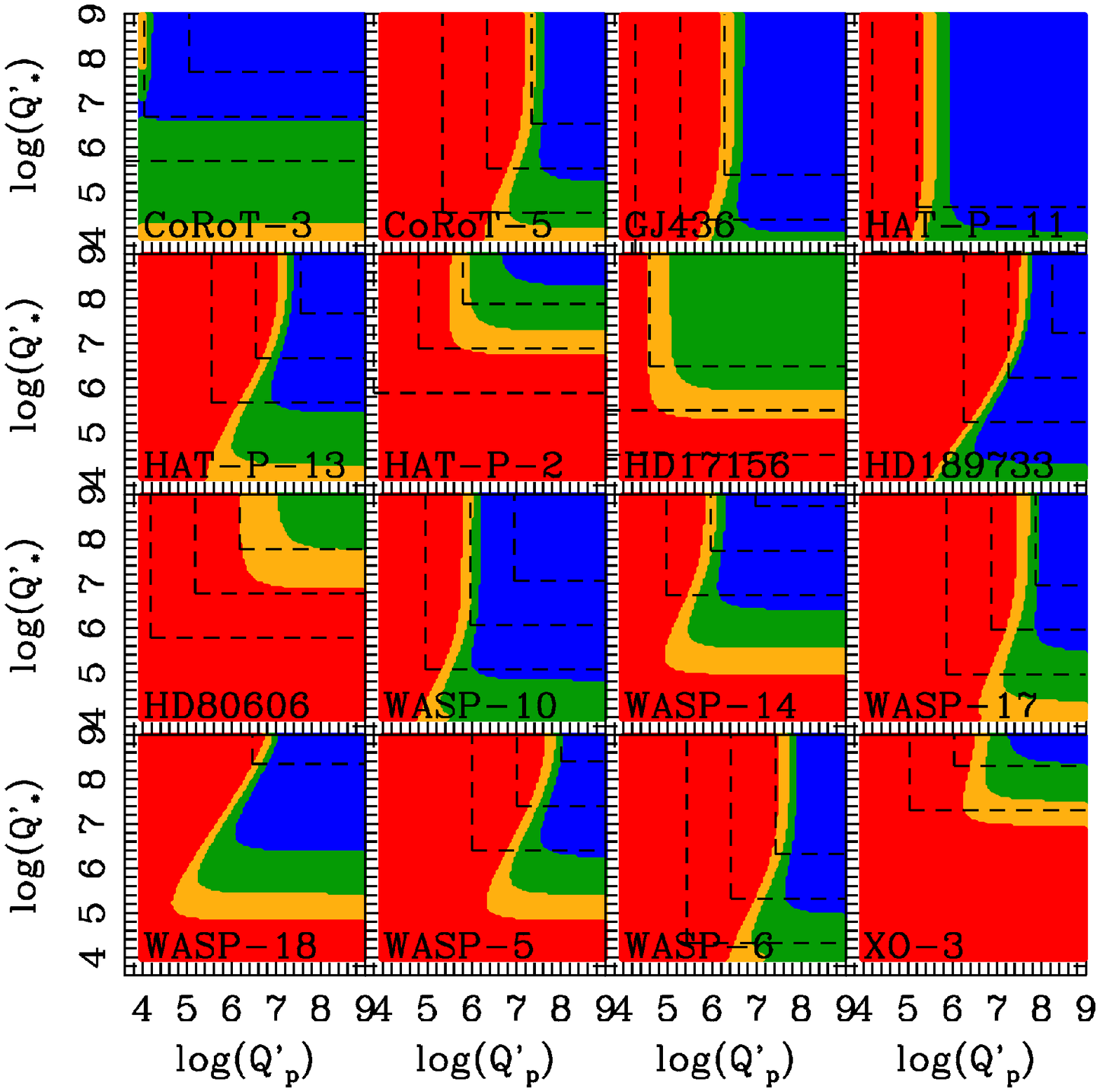}
\caption{Combinations of stellar and planetary tidal quality factors
that allow a planet to survive (i.e.,~$e<1$) for the stellar age
with uncertainties in backward integration of the tidal equations.
Magnetic braking is not included, and tidal quality factors change
as $Q'=Q'_0n_0/n$. Blue, green, orange, and red areas correspond to
a maximum zero-age semi-major axis of $0.1$, $1$, $10$, and
$\geq10\,$AU, respectively.  Also plotted are the same vertical and
horizontal dashed lines shown in Fig.~\ref{fig4}. \label{fig5}}
\end{figure}

We select systems that have solutions such that $e<1$ somewhere in
the interval $\tau_{age,\,min}\leq t \leq \tau_{age,\,max}$,
and we plot the corresponding \(Q'_{*,0}\) and \(Q'_{p,0}\) values
in Fig.~\ref{fig5}. Again, the tidal quality factors change as
$Q'=Q'_0n_0/n$, and magnetic braking is neglected. Blue, green,
orange, and red regions represent the maximum reachable ``zero-age''
semi-major axis for each system being 0.1, 1, $10\,$AU, and
$\geq10\,$AU, respectively. Vertical and horizontal dashed lines are
the same as in Fig.~\ref{fig4}, and indicate the minimum $Q'_{p,0}$
and $Q'_{*,0}$ required to have circularization and orbital decay
times (i.e., the future survival time) of $0.1$, $1$, and $10\,$Gyr,
respectively. As expected, when tidal dissipation is inefficient in
both planet and star (top right corner area of the figure), the
planets are generally expected to have stayed where they are now for
a long period of time. Lack of a significant change in its orbit,
especially eccentricity, may be consistent with the initial orbital
properties expected from planet migration scenario. For the future
survival time much longer than $10\,$Gyr, only such a solution with
little migration is allowed in the parameter space. On the other
hand, when tidal dissipation is highly efficient (bottom left
corner), a planet could have started on a wide, eccentric orbit,
which was then circularized over the lifetime of the system. The
property of this area is in good agreement with the expectation from
Kozai cycles and/or planet--planet interactions followed by
tidal dissipation. The comparison of these regions with dashed lines
imply that most planetary systems have a wide range of allowed
parameter space for $(Q'_{*,0},\,Q'_{p,0})$ which is consistent with
having started on a wide, eccentric orbit, and with comfortable
future survival times $\sim 1-10\,$Gyr. Thus, our results agree with
the suggestion first made by \cite{Jackson08}, and show that there
is a broad parameter space which supports the tidal circularization
scenario as the dominant origin of close-in planets.

When we demand that the future survival time must be comparable to,
or slightly longer than the estimated stellar age, Fig.~\ref{fig5}
implies an interesting trend. In the stellar-dissipation dominated
region, planets tend to have an initial orbit which is similar to
the current one, while in the planetary-dissipation dominated
region, planets tend to have an initially highly eccentric and wide
orbit. This is clear by comparing these two regions (below and above
the diagonal line that can be drawn by connecting the vertices of
vertical and horizontal dashed lines) at around their survival
times. Most systems in the figure except WASP-10, WASP-14, WASP-18,
and WASP-6 have the estimated age of $\sim 1-10\,$Gyr, while
WASP-10, WASP-14, and WASP-18 have $\sim 0.1-1\,$Gyr, and WASP-6 has
$> 10\,$Gyr. Investigating the corresponding regions, we find that
the maximum zero-age semi-major axis can be very large (upto
$>10\,$AU) in the planetary-dissipation dominated region. On the
other hand, such a solution is less likely in the
stellar-dissipation dominated region, and the area with little
change in orbital radii tends to occupy the largest parameter space.


\begin{figure}
\plotone{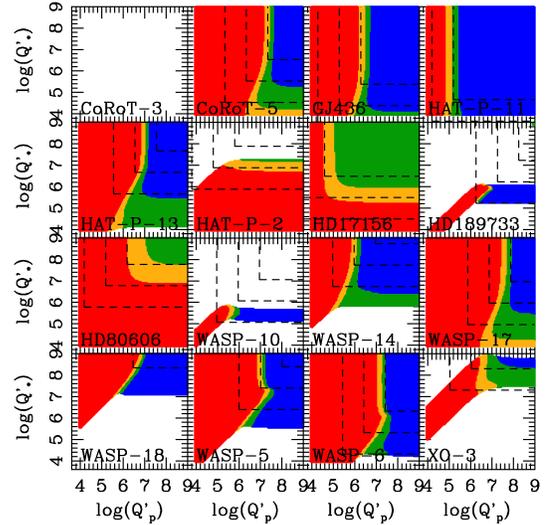}
\caption{Same as Fig.~\ref{fig5}, but with the effects of magnetic
braking included. Systems that are affected the most by magnetic
braking have the rapidly rotating star, and/or a relatively high
mass ratio. For CoRoT-3, we don't get any solutions. \label{fig6}}
\end{figure}

Fig.~\ref{fig6} presents similar results with magnetic braking
effects included. As we can see, some systems are much less affected
by magnetic braking than others. Our results show that, when either
the stellar spin is slow, or the mass ratio is very low, the
evolution is less affected by magnetic braking. Clearly affected
systems include CoRoT-3, HAT-P-2, HD~189733, WASP-10, WASP-14,
WASP-18, WASP-5, and XO-3, for which the average stellar spin period
is $\simeq 8\,$days and the average mass ratio is $\simeq 0.006$,
while the corresponding values are about $36\,$days and $0.001$ for
the others.

%

\subsection{Stellar Obliquity Evolution}
\begin{figure}
\plotone{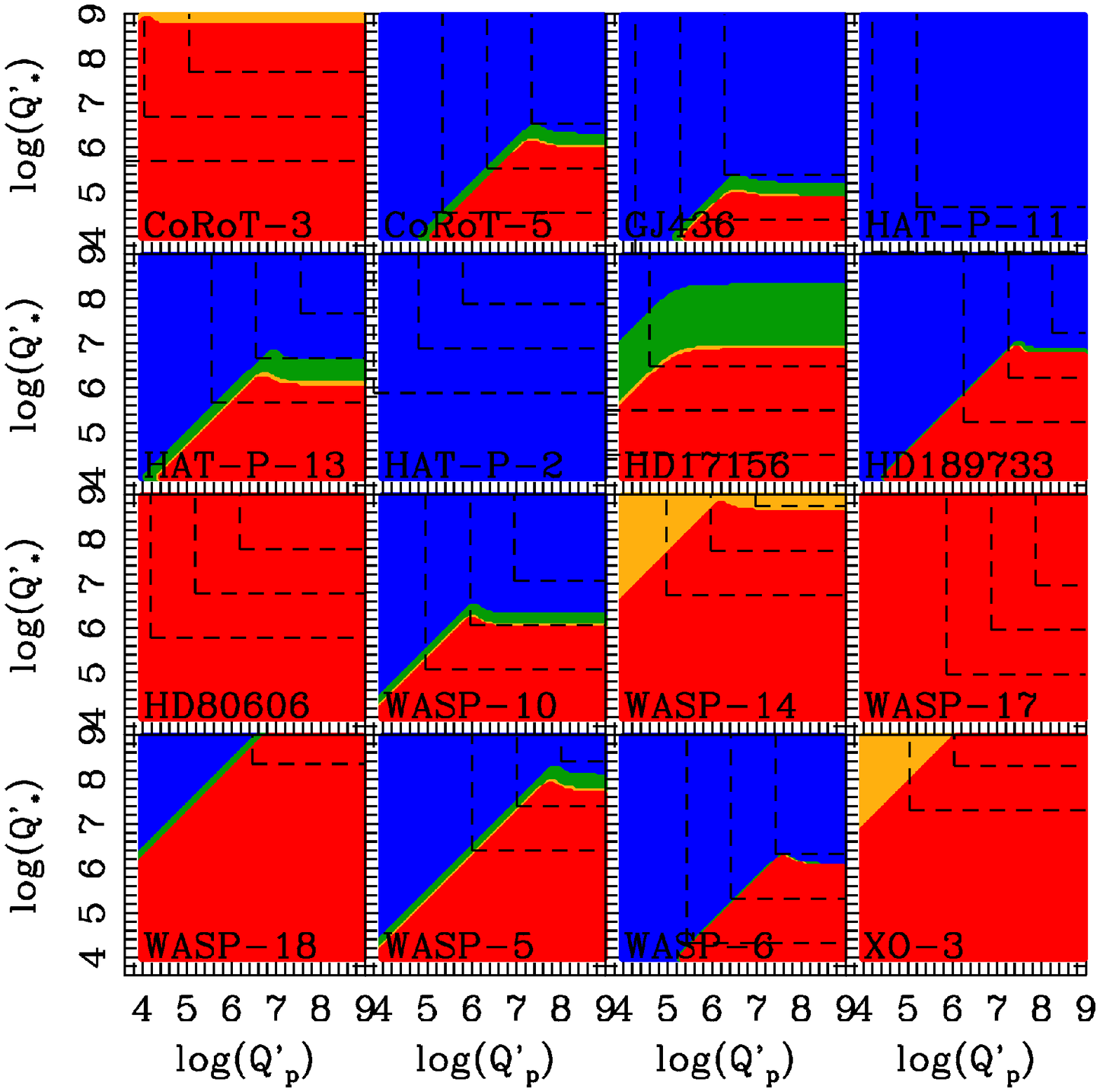}

\caption{Combinations of stellar and planetary tidal quality factors
that allow a planet to survive (i.e.,~$e<1$) for the stellar age
with uncertainties in backward integrations of the tidal equations.
Magnetic braking is not included, and tidal quality factors are
scaled as $Q'=Q'_0n_0/n$. Blue, green, orange, and red areas
correspond to a maximum zero-age stellar obliquity of $5$, $20$,
$40$, and $\geq40\,$degrees, respectively. Also plotted for
reference are the same vertical and horizontal dashed lines shown in
Fig.~\ref{fig4}. \label{fig7}}

\end{figure}

\begin{figure}
\plotone{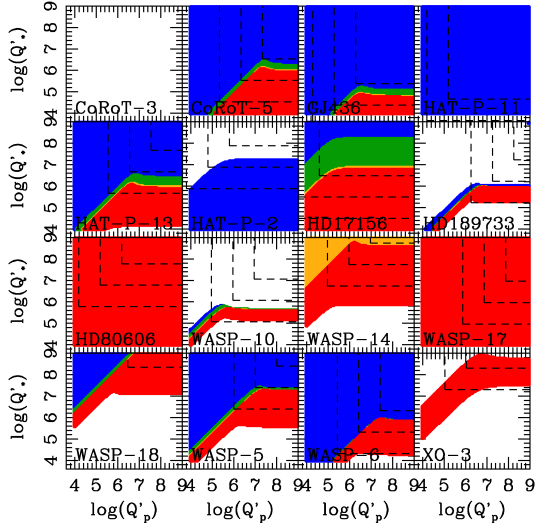}

\caption{Same as Fig.~\ref{fig7}, but with magnetic braking
included. Systems which are affected the most have the rapidly
rotating star, and/or a relatively high mass ratio. For CoRoT-3, we
don't get any solutions. \label{fig8}}

\end{figure}

Fig.~\ref{fig7}, and \ref{fig8} show similar plots to
Fig.~\ref{fig5}, and \ref{fig6} for the stellar obliquity. Blue,
green, orange, and red areas correspond to a maximum possible
zero-age obliquity of $5^{\circ}$, $20^{\circ}$, $40^{\circ}$, or
$\geq40^{\circ}$. Again, also plotted are the same vertical and
horizontal lines as in Fig.~\ref{fig4}.

In Fig.~\ref{fig7}, and \ref{fig8}, CoRoT-3, HAT-P-2, HD~17156,
HD~80606, WASP-14, WASP-17, WASP-18, WASP-5, WASP-6, and XO-3 have
RM measurements, and thus known projected stellar obliquities.
Among them, CoRoT-3, HD~80606, WASP-14, WASP-17, and XO-3 have a
significant misalignment ($>20^{\circ}$). Naturally, the maximum
possible misalignment at zero age can be very large
($\geq40^{\circ}$) for these systems with almost any values of
$(Q'_{*},\,Q'_{p})$. On the other hand, for systems with small
measured projected obliquities (HAT-P-2, HD~17156, WASP-18, WASP-5,
and WASP-6), or systems with no RM measurements (CoRoT-5, GJ 436,
HAT-P-11, HAT-P-13, HD~189733, and WASP-10), we find either little
or large change in obliquity over the stellar age.

Consider a few specific cases. For HAT-P-2 and HAT-P-11, there are
no solutions for obliquity larger than $5^{\circ}$. This indicates
that, for HAT-P-2, the ``zero-age'' obliquity was likely similar to
the current nominal value $\lambda=1.2^{\circ}$. However, since
HAT-P-2's measured obliquity has a very large uncertainty
($\lambda=1.2^{\circ}\pm13.4^{\circ}$), it is possible that the
actual present value is much larger. Note that even if the stellar
obliquity turns out to be small, HAT-P-2 is still likely to have the
scattering/Kozai-cycle origin. This is partly because of its high
eccentricity ($e\sim0.52$), and partly because of a large range of
tidal quality factors that allow a much wider orbit in the past (see
Fig.~\ref{fig5}, and \ref{fig6}). For HAT-P-11, there are no RM
measurements so far, but the system shows a similar result to
HAT-P-2 with $\epsilon_{*,0}=2^{\circ}$. If the current obliquity
turns out to be $\lesssim2^{\circ}$, our plot indicates that
HAT-P-11 would not have had a large obliquity in the past. This in
turn indicates that, if HAT-P-11 initially had a large stellar
obliquity, we should be able to see a clearly misaligned orbit
through future observations, independent of the actual tidal quality
factors for the system. The orbital eccentricity of the planet is
$e=0.198\pm0.046$, which could have been produced either via
planet--disk, or planet--companion interactions. Interestingly,
Fig.~\ref{fig5} and \ref{fig6} show that the orbital radius of
HAT-P-11~b is unlikely to have been changed significantly via tidal
evolution, unless the tidal dissipation inside the planet is rather
efficient ($Q'_p<10^{6}$). If the tidal dissipation is inefficient,
our result suggests that HAT-P-11~b is likely to have the migration
origin.
We have to wait for future observations to estimate whether the
planet is likely formed via disk migration, or tidal circularization
of a highly eccentric orbit\footnote{As we were writing up this
draft, two groups released the results that HAT-P-11 is a highly inclined system. 
The sky-projected stellar obliquity is $\lambda=103^{+26}_{-10}\,$degrees by 
\cite{Winn10HATP11ap}, and $103^{+23}_{-19}\,$degrees by \cite{Hirano10ap}. 
Thus, the system is likely to have the scattering/Kozai-cycle origin, rather than the migration
one.}.

For all the other systems, the zero-age obliquity can be as high as
$\geq40^{\circ}$ if stellar tidal quality factors are relatively
small. More specifically, such solutions are allowed in the
stellar-dissipation dominated region with $\tau_e\simeq \tau_a$.
The stellar obliquity generally damps on a similar timescale to the
orbital decay \citep[see, e.g., LWC09,][as well as
Section~4.1]{Barker09}. However, this plot demonstrates that the
stellar obliquity could be damped from high ($\gtrsim 40^{\circ}$)
to low ($\sim2^{\circ}$) values within the current stellar age.

Note that, these small stellar tidal quality factors which allow the
fast damping of stellar obliquities, also lead to relatively short
survival times for planets. By comparing these zero-age high
obliquity regions with the horizontal dashed lines, we find that
these tidal quality factors lead to survival times at most
comparable to the current stellar age. Or in other words, relatively
small parameter spaces are allowed for these high zero-age
obliquities.
For example, when we demand that the expected survival time in
forward tidal evolution must be comparable to or larger than the
stellar age, we find that the stellar tidal quality factor for
CoRoT-5 must be $Q'_{*}\gtrsim10^{6}$. This includes a small region
of the red, large ``zero-age'' stellar obliquity area in
Fig.~\ref{fig7}. However, if $Q'_{*}\gtrsim10^{7}$, such an area
disappears. In the latter case, if the observations find a small
stellar obliquity for CoRoT-5, it is unlikely that the stellar
obliquity was much larger in the past. In some cases, such a region
with a high zero-age obliquity and a comfortable future survival
time may not even exist. For HAT-P-13, a similar comparison shows
that the stellar tidal quality factor must be $Q'_{*}\gtrsim10^{7}$
for the survival time to be comparable to or larger than the stellar
age. This corresponds to the blue, small ``zero-age'' stellar
obliquity area in Fig.~\ref{fig7}, and thus we expect that the
stellar obliquity of HAT-P-13 could not have been changed much due
to tidal evolution (i.e.,~similar to HAT-P-2 or HAT-P-11).
Interestingly, HAT-P-13 is the only system in our sample that has an
additional planetary companion. Thus, the system may be an example
of a close-in planet formed via migration scenario. If this is the
case, we predict the future observations will find a small stellar
obliquity for HAT-P-13 \footnote{Indeed, a recent observation found
a well-aligned orbit for HAT-P-13b with
$\lambda=-1.9\pm8.6\,$degrees \citep{Winn10HATP13}.}.

In short, by comparing Fig.~\ref{fig5} with Fig.~\ref{fig7}, we find
that the evolution history is largely divided into two cases,
depending on the relative efficiency of energy dissipation inside
the star and the planet. In the planetary-dissipation dominated
region, the planets could have had a wide, eccentric orbit in the
past, and the stellar obliquity damps on a similar timescale to the
orbital decay. 
The initial conditions implied in this region are consistent with those 
expected from the scattering/Kozai-cycle origin of the close-in planets. 
On the other hand, in the stellar-dissipation
dominated region, the system could have had a large stellar
obliquity in the past, although the initial orbital radius and
eccentricity are likely similar to the current values, as 
planet migration scenario would suggest.
The further inspection of these figures show that a unique evolution
is possible in the transition region of these two, where the
dissipation effects due to a planet and a star are comparable.
There, the system could have had a wide, eccentric orbit, as well as
a large stellar obliquity in the past.
Our results also imply that if the currently observed stellar
obliquity distribution is due to Kozai migration as suggested by
\cite{Triaud10ap}, then most exoplanetary systems have
planetary-dissipation dominated tidal interactions
(i.e.,~$\tau_e<\tau_a$). If $\tau_e\simeq \tau_a$ for most planetary
systems, the current obliquity distribution should be very different
from the initial one, and the dynamical history prior to the tidal
dissipation should be wiped out.

Recently, \cite{Winn10} suggested that the stellar obliquity is
preferentially large for hot stars with effective temperatures
$T_{\rm eff}>6250\,$K, and proposed that such a trend can be
explained because photospheres of cool stars can realign with the
orbits due to tidal dissipation in the convective zones, without
affecting the orbital decay. Our results suggest that yet another
possibility might be that the evolution of the systems with hot and
cool stars correspond to slow and fast obliquity damping regions,
respectively. If that is the case, the stellar obliquity may stay
similar to the original value for the planetary system with a hot
star, because tidal dissipation is dominated by the planet
(i.e.,~$\tau_e<\tau_a$), while the obliquity may be damped to a
small value for the system with a cool star, because tidal
dissipation is either dominated by the star (i.e.,~$\tau_e\simeq
\tau_a$), or the star and the planet have the comparable dissipation
effects.
In our samples of eccentric systems, CoRoT-5, GJ 436, HAT-P-11,
HAT-P-13, HD~17156, HD~189733, HD~80606, WASP-10, WASP-5, and WASP-6
have the effective temperatures less than $6250\,$K. For the range
of tidal quality factors we use, we don't see any trend that these
cool systems prefer $\tau_e\simeq \tau_a$. One possibly interesting
example is HD~17156, for which the large parameter space is allowed
for the stellar-dissipation dominated case ($\tau_e\simeq \tau_a$).

\section{Different Scales for Tidal Quality Factors}
So far, we have focused on a scaling of $Q\propto1/n$. However, this
scaling is not appropriate unless a close-in planet system reaches a
true dual synchronization of $n=\omega_*=\omega_p$. This cannot be
reached unless both eccentricity and obliquities are zero, or some
extra torques are acting on the system. For a planetary spin,
although the true synchronization with the orbit normal does not
happen until the final spiral-in of the planet, the
pseudo-synchronization reaches quickly (see e.g.,~Section~4.1 and
4.2). On the other hand, a stellar spin changes on a similar
timescale to the orbital decay, and thus it's reasonable, for most
systems, to assume that the stellar spin is far from
synchronization. As we have seen in Section~3.1, the semi-diurnal
tide with the forcing frequency of $|2\omega-2n|$ dominates the
energy dissipation before the spin-orbit synchronization, while the
annual tide with $|2\omega-n|$ takes over once the synchronization
approaches. Therefore, in this section, we scale the planetary and
stellar tidal quality factors as $Q_p\propto1/|2\omega_p-n|$ and
$Q_*\propto1/|2\omega_*-2n|$, respectively, and investigate the
differences in future and past evolution from the results of the
scaling $Q\propto1/n$.

\begin{figure}
\plotone{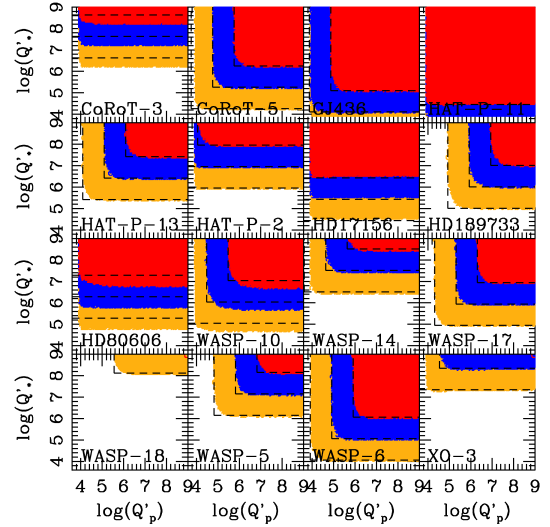}
\caption{Same as Fig.~\ref{fig4}, but with different scalings for tidal quality factors.
\label{StanQsQp2_fw}}

\end{figure}

In Fig.~\ref{StanQsQp2_fw}, we repeat the similar tidal evolution
forward in time as Fig.~\ref{fig4} with the initial tidal quality
factors ranging over the interval $10^4\leq Q'_{0}\leq 10^9$. The
figure has a similar general trend to Fig.~\ref{fig4}, but the
larger parameter space is allowed. The vertical and horizontal lines
are obtained by rescaling Eq.~\ref{dedtapprox} and \ref{dadtapprox}
by $n_0/|2\omega_{p,0}-n_0|$ and $n_0/|2\omega_{*,0}-n_0|$,
respectively. The agreement with the integration of the complete
tidal equations is pretty good.

\begin{figure}
\plotone{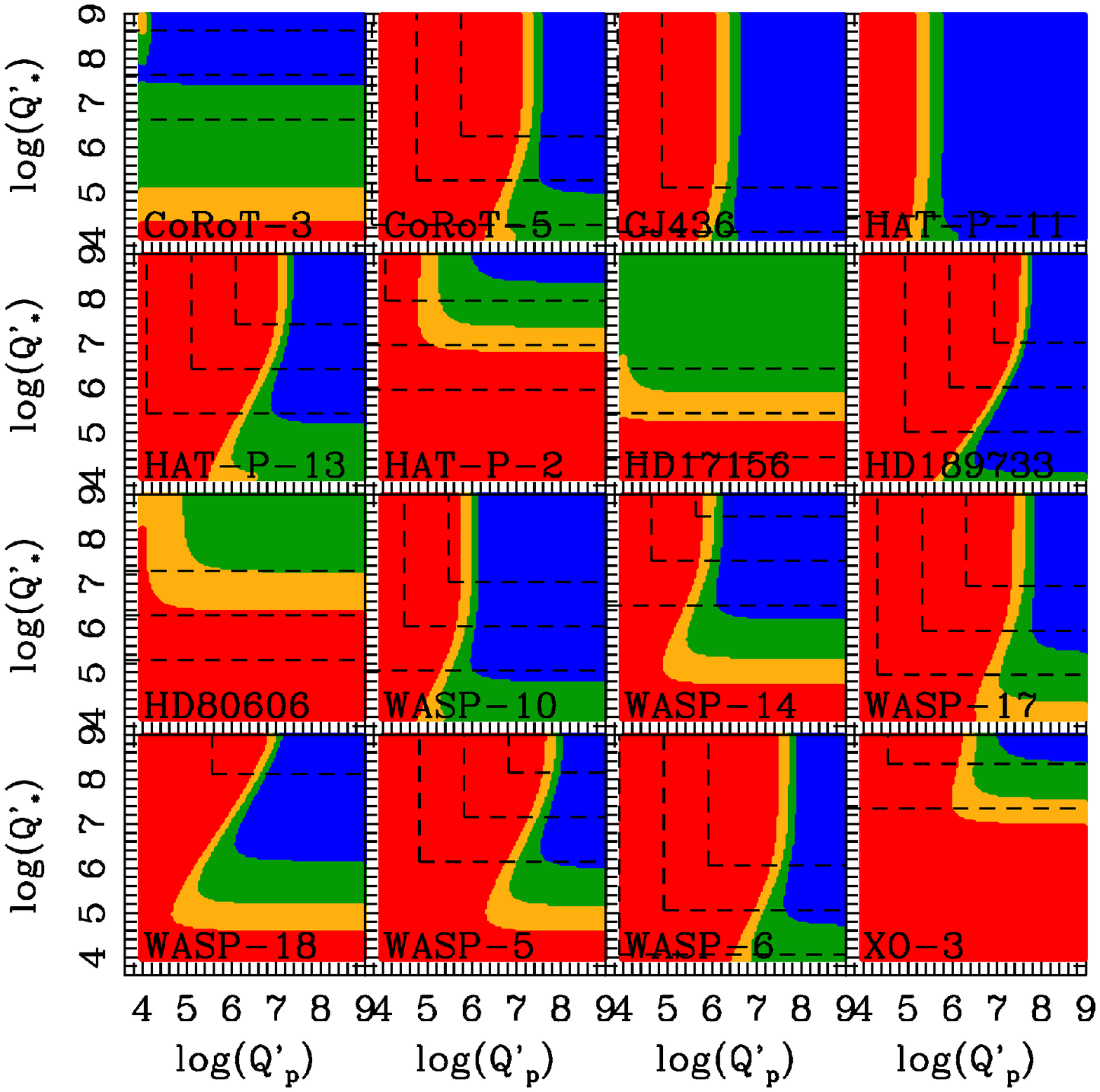}
\caption{Same as Fig.~\ref{fig5}, but with different scalings for tidal quality factors.
\label{StanQsQp2_bw}}
\end{figure}

\begin{figure}
\plotone{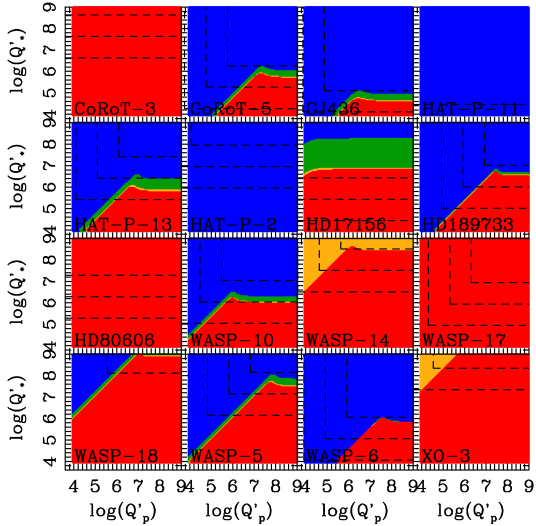}
\caption{Same as Fig.~\ref{fig7}, but with different scalings for tidal quality factors.
\label{StanQsQp2_bw_obliq}}
\end{figure}

Similarly, Fig.~\ref{StanQsQp2_bw} and \ref{StanQsQp2_bw_obliq} show
the corresponding results to Fig.~\ref{fig5} and \ref{fig7},
respectively, with the modified Q scalings. Both of these figures
have a good agreement with the $Q\propto1/n$ case.

%

\section{Discussion and Conclusions}

Close-in planets may be formed via planet migration in a disk, or
via tidal circularization of a highly eccentric orbit following
planet--planet scattering or other gravitational interactions with a
companion. There is strong observational support for the tidal
circularization scenario. First, the current observations exhibit
different orbital distributions for apparently single- and
multiple-planet systems \citep{Wright09}. The so-called ``3-day
pileup'' and a jump in planetary abundance beyond 1 AU are not seen
among multi-planet systems. It is difficult to explain such a
difference as a result of planet migration, since migration in the
disk without strong planet--planet interactions is expected to lead
to similar orbital distributions for both single and multiple planet
systems. On the other hand, strong gravitational interactions
between planets and/or stellar companions can disrupt the
multiple-planet systems by ejecting planets or scattering them far
away from one another. Thus, from such a scenario, it's expected
that systems which have gone through violent dynamical interactions
have only one planet close to the central star, while systems which
did not experience such an event retain multiple planets close-by.
Second, the observed distribution of stellar obliquities matches
very well with the expectation from Kozai migration
\citep{Triaud10ap}. Third, the observed inner edge of the orbital
distribution is not at the Roche limit $a_{\rm R}$, which would be
expected from planet migration, but rather at $2a_{\rm R}$, which is
naturally explained by circularization of a highly eccentric orbit
while conserving total angular momentum \citep{Ford06}.

In this paper, we have further explored the possibility of forming
close-in planets via tidal circularization of an eccentric orbit by
studying the past and future evolution of eccentric transiting
planets. We considered a broad range of tidal quality factors, $\sim
10^4-10^9$, consistent with our limited theoretical understanding of
tidal dissipation in both stars and planets. We have used a tidal
model where $Q'\propto 1/(\sigma\Delta t)$ as the simplest
representation because of the unknown character of tidal dissipation
in stars and gaseous planets. Our choice is consistent with the
constant time lag model \citep{Hut81} that is derived as a
quadrupolar approximation of the tidal potential. Although different
tidal models can change the time scales of the various processes,
the qualitative properties of the evolution should be the same as 
other models.
Another caveat here is that, for these close-in planets, the ratio
$R_*/a$ is relatively large and often $\sim0.1$ (e.g.,~see
Fig.~\ref{forwardQL09ae}). Thus, the higher-order terms in the tidal
potential can be important, as they are for Mars' satellites Phobos
and Deimos \citep[e.g.,][]{Bills05}.

In Section~2 we investigated the effects of (known or unknown)
planetary/stellar companions on the evolution of observed close-in
planets. Comparing secular timescales with GR precession timescales,
we showed that the current and future evolution of most close-in
planets is unlikely to be strongly affected by such companions. In
Section~4.1 we re-examined the tidal stability of transiting
systems, and confirmed that the majority of close-in planets are
Darwin-unstable. The exceptions include CoRoT-3, CoRoT-6, HD~80606,
and WASP-7 which are Darwin-stable within the current observational
uncertainties.
We also found that borderline cases like HAT-P-2
and WASP-10 are most likely Darwin-unstable, but could be Darwin-stable
within the uncertainties.

For clearly Darwin unstable systems (with $L_{\rm tot}\ll L_{\rm
crit}$), there are two possible evolutionary paths. When the tidal
dissipation in the star dominates the evolution, all parameters but
the planetary spin evolve on a similar timescale
($\tau_{e}\sim\tau_{a}\sim\tau_{\epsilon_{*}}\sim\tau_{\omega_{*}}$).
On the other hand, when the tidal dissipation in the planet is
non-negligible, the orbit tends to get circularized before the
planet spirals all the way to the Roche limit
($\tau_{e}<\tau_{a}\sim\tau_{\epsilon_{*}}\sim\tau_{\omega_{*}}$).
Although it is nontrivial to determine which evolutionary path each
system will take without knowing the efficiencies of tidal
dissipation in the star and in the planet, there are some
indications that dissipation in the planet dominates
($\tau_{e}<\tau_{a}$) for most systems. First, there is clear
evidence of eccentricity damping within $\sim0.1\,$AU. Most observed
close-in planets are on nearly circular orbits, which are
well-described by the traditional exponential eccentricity damping
approximation. Second, our results in Section~5.2 suggest that we
need to assume $\tau_{e}<\tau_{a}$ for most systems in order to
explain the current obliquity distribution through Kozai migration.
With stellar-dissipation dominated cases ($\tau_{e}\sim\tau_{a}$),
we expect the distribution of current stellar obliquity would be
significantly different from the one expected from Kozai migration.

In Section~4.3, we showed that the lifetime of the planetary systems
is largely determined by tidal dissipation in the star, while the
circularization time is largely determined by dissipation in the
planet. Also, we confirmed the results by \cite{Barker09} and showed
that magnetic braking does not have a large effect for the future
evolution except for systems with a rapidly rotating star. The
minimum stellar tidal quality factor that allows a planet to survive
for a certain age is similar for the cases with and without magnetic
braking (see Fig.~\ref{fig4} and \ref{forwardmbQL09all}).

In Section~5, we found that, generally speaking, the evolution
history in the stellar-dissipation dominated tidal evolution is
consistent with that expected from the planet migration origin of
the close-in planets, with little change in semi-major axis and
eccentricity over the stellar age. On the other hand, the evolution
history in the planetary-dissipation dominated evolution is
consistent with that expected from scattering/Kozai-cycle origin,
with initial orbits being wide and eccentric.
The latter case agrees with the results in \cite{Jackson08}. 
We also showed that, when the effects of tidal dissipation in the star is 
comparable to, or larger than that in the planet, 
the stellar obliquities could be
damped from high $\gtrsim40^{\circ}$ to low $\sim2^{\circ}$ values
within the currently observed stellar ages 
(see~Fig.~\ref{fig7}, and \ref{fig8}).

Overall, our results for the tidal evolution of eccentric transiting
planets is consistent with the formation path that involves
circularization of an initially eccentric orbit.  The distribution
of system parameters seems to imply that this mechanism dominates
for close, single planet systems, but multiplanet systems are more
consistent with the migration to account for the close members.

\acknowledgements This work was supported by an Astronomy Center for
Theory and Computation Prize Fellowship (to S.~M.) at the University
of Maryland, and by NSF Grant AST--0507727 (to F.~A.~R.) at
Northwestern University. Partial support was also provided by the
Center for Interdisciplinary Exploration and Research in
Astrophysics (CIERA) at Northwestern University.
S.~J.~Peale is supported in part by NASA under grant NNG06GF42G in
the Origins of Solar Systems Program. S.~M. also thanks the support
from the Kavli Institute for Theoretical Physics at UC Santa Barbara
while part of this work was done. We thank Benjamin Levrard, Giles
Chabrier, Brian Jackson, Amaury Triaud, Dong Lai, Yoram Lithwick,
Smadar Naoz for useful discussions, and an anonymous referee for
valuable comments on the manuscript. S.~M. thanks Genya Takeda for
making the time comparison code used in Fig.~\ref{fig0} and
\ref{companionall} available.
\bibliographystyle{apj}
\bibliography{REF}
\clearpage
\LongTables
\begin{landscape}
\begin{deluxetable}{lllllllllll}
\tabletypesize{\scriptsize}
\tablewidth{0pt}
\tablecaption{Data are taken from {\tt http://exoplanet.eu/}.  \label{tab1}}
\setlength{\tabcolsep}{0.01in}
%
%
\tablecolumns{11} \tablehead{ \colhead{Planet Name} &  \colhead{$M_{\rm p}$} & \colhead{$R_{\rm p}$} & \colhead{$a$} &
\colhead{e} & \colhead{$M_*$} & \colhead{$R_*$} & \colhead{$v\sin i$} & \colhead{$\lambda$} & \colhead{Age} & \colhead{References} \\
\colhead{} &  \colhead{[$M_J$]} & \colhead{[$R_J$]} & \colhead{[AU]} &
\colhead{} & \colhead{[$M_{\odot}$]} & \colhead{[$R_{\odot}$]} & \colhead{[km/s]} & \colhead{[degrees]} & \colhead{[Gyr]} & \colhead{}
}

\startdata
CoRoT-1~b (G0V) & $1.03^{+0.12}_{-0.12}$ & $1.49^{+0.08}_{-0.08}$ &  $0.0254^{+0.0004}_{-0.0004}$ &  0 (fixed) & $0.95^{+0.15}_{-0.15}$ &  $1.11^{+0.05}_{-0.05}$ & $5.2^{+1.0}_{-1.0}$ & $-77^{+11}_{-11}$ & & Barge08, Pont10 \\
CoRoT-2~b (G7V) & $3.31^{+0.16}_{-0.16}$ &  $1.465^{+0.029}_{-0.029}$ &  $0.0281^{+0.0009}_{-0.0009}$ &  0 (fixed) & $0.97^{+0.06}_{-0.06}$ &  $0.902^{+0.018}_{-0.018}$ & $11.85^{+0.50}_{-0.50}$ &  $7.2^{+4.5}_{-4.5}$ & $\sim$0.2-4 & Alonso08, Bouchy08 \\
CoRoT-3~b (F3V) & $21.23^{+0.82}_{-0.59}$ &  $0.9934^{+0.058}_{-0.058}$ &  $0.05694^{+0.00096}_{-0.00079}$ &  $0.008^{+0.015}_{-0.005}$ & $1.359^{+0.059}_{-0.043}$ & $1.540^{+0.083}_{-0.078}$ & $17.0^{+1.0}_{-1.0}$ & $-37.6^{+22.3}_{-10.0}$ & 1.6-2.8 & Triaud09, Deleuil08 \\
CoRoT-4~b (F0V) & $0.72^{+0.08}_{-0.08}$ &  $1.190^{+0.06}_{-0.05}$ & $0.09^{+0.001}_{-0.001}$ & $0^{+0.1}_{-0.1}$ & $1.16^{+0.03}_{-0.02}$ & $1.17^{+0.01}_{-0.03}$ & $6.4^{+1.0}_{-1.0}$ & &  $1^{+1.0}_{-0.3}$ & Moutou08 \\
CoRoT-5~b (F9V) & $0.467^{+0.047}_{-0.024}$ &  $1.388^{+0.046}_{-0.047}$ & $0.04947^{+0.00026}_{-0.00029}$ &   $0.09^{+0.09}_{-0.04}$ &  $1.00^{+0.02}_{-0.02}$ & $1.186^{+0.04}_{-0.04}$ & $1^{+1}_{-1}$ & & 5.5-8.3 & Rauer09 \\ 
CoRoT-6~b (F9V) & $2.96^{+0.34}_{-0.34}$ &  $1.166^{+0.035}_{-0.035}$ & $0.0855^{+0.0015}_{-0.015}$ &  $<0.1$ &  $1.05^{+0.05}_{-0.05}$ & $1.025^{+0.026}_{-0.026}$ & $7.6^{+1.0}_{-1.0}$ & & 2.5-3.3 & Fridlund10 \\
CoRoT-7~b (G9V) & $0.0151^{+0.0025}_{-0.0025}$ & $0.15^{+0.008}_{-0.008}$ & $0.0172^{+0.00029}_{-0.00029}$ &  0 & $0.93^{+0.03}_{-0.03}$ & $0.87^{+0.04}_{-0.04}$ &  & &  1.2-2.3 & Queloz09 \\
CoRoT-7~c (G9V) & $0.0264^{+0.0028}_{-0.0028}$ & & $0.046$ & 0 & & & & & & Queloz09 \\
GJ 1214~b (M) & $0.0204$ & $0.239$ & $0.0143$ & $<0.27$ & $0.157^{+0.019}_{-0.019}$ & $0.211^{+0.0097}_{-0.0097}$ & $<2.0$ & & 3-10 & Charbonneau09 \\
GJ 436~b (M2.5) &  $0.0729^{+0.0025}_{-0.0025}$ &  $0.3767^{+0.0082}_{-0.0092}$ & $0.02872^{+0.00029}_{-0.00026}$ & $0.14^{+0.01}_{-0.01}$ &  $0.452^{+0.014}_{-0.012}$ & $0.464^{+0.009}_{-0.011}$ & $0.52^{+0.05}_{-0.05}$ & & $6.0^{+4.0}_{-5.0}$ & TWC08, LWC09 \\
HAT-P-1~b (G0V) & $0.532^{+0.030}_{-0.030}$ & $1.242^{+0.053}_{-0.053}$ & $0.0553^{+0.0012}_{-0.0013}$ & $<0.067$ & $1.133^{+0.075}_{-0.079}$ & $1.135^{+0.048}_{-0.048}$ & $3.75^{+0.58}_{-0.58}$ &  $3.7^{+2.1}_{-2.1}$ & $2.7^{+2.5}_{-2.0}$ & TWC08, Johnson08 \\
HAT-P-11~b (K4) & $0.081^{+0.009}_{-0.009}$ &  $0.422^{+0.014}_{-0.014}$ &  $0.0530^{+0.0002}_{-0.0008}$ & $0.198^{+0.046}_{-0.046}$ & $0.81^{+0.02}_{-0.03}$ &  $0.75^{+0.02}_{-0.02}$ &  $1.5^{+1.5}_{-1.5}$ & & $6.5^{+5.9}_{-4.1}$ & Backos10 \\
HAT-P-12~b (K4) &  $0.211^{+0.012}_{-0.012}$ & $0.959^{+0.029}_{-0.021}$ & $0.0384^{+0.0003}_{-0.0003}$ & 0 & $0.733^{+0.018}_{-0.018}$ & $0.701^{+0.02}_{-0.01}$ & $0.5^{+0.4}_{-0.4}$ & & $2.5^{+2.0}_{-2.0}$ & Hartman09 \\
HAT-P-13~b (G4) & $0.853^{+0.029}_{-0.046}$ & $1.281^{+0.079}_{-0.079}$ &  $0.0427^{+0.0006}_{-0.0012}$ & $0.021^{+0.009}_{-0.009}$ & $1.22^{+0.05}_{-0.10}$ &  $1.56^{+0.08}_{-0.08}$ &  $2.9^{+1.0}_{-1.0}$ & & $5.0^{+2.5}_{-0.7}$ & Bakos09a \\
HAT-P-13~c (G4) & $15.2^{+1.0}_{-1.0}$ & & $1.188^{+0.018}_{-0.033}$ & $0.691^{+0.018}_{-0.018}$ & & & & & & Bakos09a \\
HAT-P-2~b (F8) & $9.09^{+0.24}_{-0.24}$ & $1.157^{+0.073}_{-0.062}$ & $0.06878^{+0.00068}_{-0.00068}$ & $0.5171^{+0.0033}_{-0.0033}$ & $1.36^{+0.04}_{-0.04}$ &  $1.64^{+0.09}_{-0.08}$ &  $20.8^{+0.3}_{-0.3}$ &  $1.2^{+13.4}_{-13.4}$ &  $2.6^{+0.5}_{-0.5}$ & P\'{a}l10, Winn07 \\
HAT-P-3~b (K) & $0.596^{+0.024}_{-0.026}$ & $0.899^{+0.043}_{-0.049}$ & $0.03882^{+0.00060}_{-0.00077}$ & 0 & $0.928^{+0.044}_{-0.054}$ & $0.833^{+0.034}_{-0.044}$ & $0.5^{+0.5}_{-0.5}$ & & $1.5^{+5.4}_{-1.4}$ & TWC08, Torres07 \\
HAT-P-4~b (F) & $0.68^{+0.04}_{-0.04}$ & $1.27^{+0.05}_{-0.05}$ & $0.0446^{+0.0012}_{-0.012}$ &  0 & $1.26^{+0.06}_{-0.14}$ & $1.59^{+0.07}_{-0.07}$ & $5.5^{+0.5}_{-0.5}$ & & $4.2^{+2.6}_{-0.6}$ & Kov\'{a}cs07 \\
HAT-P-5~b (G) & $1.06^{+0.11}_{-0.11}$ & $1.26^{+0.05}_{-0.05}$ & $0.04075^{+0.00076}_{-0.00076}$ & 0 & $1.160^{+0.062}_{-0.062}$ & $1.167^{+0.049}_{-0.049}$ & $2.6^{+1.5}_{-1.5}$ & & $2.6^{+1.8}_{-1.8}$ & Bakos07 \\
HAT-P-6~b (F8) & $1.057^{+0.119}_{-0.119}$ & $1.330^{+0.061}_{-0.061}$ & $0.05235^{+0.00087}_{-0.00087}$ & 0 (fixed) & $1.29^{+0.06}_{-0.06}$ &  $1.46^{+0.06}_{-0.06}$ &  $8.7^{+1.0}_{-1.0}$ & & $2.3^{+0.5}_{-0.7}$ & Noyes08 \\ 
HAT-P-7~b (F6V)& $1.776^{+0.077}_{-0.049}$ & $1.363^{+0.195}_{-0.087}$ & $0.0377^{+0.0005}_{-0.0005}$ &  0 (fixed) & $1.47^{+0.08}_{-0.05}$ &  $1.84^{+0.23}_{-0.11}$ &  $3.8^{+0.5}_{-0.5}$ & $182.5^{+9.4}_{-9.4}$ & $2.2^{+1.0}_{-1.0}$ & 
P\'{a}l08, Winn09c \\
HAT-P-8~b (F) & $1.52^{+0.18}_{-0.16}$ &  $1.50^{+0.08}_{-0.06}$ & $0.0487^{+0.0026}_{-0.0026}$ &  0 (fixed) & $1.28^{+0.04}_{-0.04}$ &  $1.58^{+0.08}_{-0.06}$ &  $11.5^{+0.5}_{-0.5}$ & & $3.4^{+1.0}_{-1.0}$ & Latham09 \\
HAT-P-9~b (F) & $0.78^{+0.09}_{-0.09}$ & $1.40^{+0.06}_{-0.06}$ &  $0.053^{+0.002}_{-0.002}$ &  0 (fixed) & $1.28^{+0.13}_{-0.13}$ & $1.32^{+0.07}_{-0.07}$ &  $11.9^{+1.0}_{-1.0}$ & & $1.6^{+1.8}_{-1.4}$ & Shporer09 \\
HD~149026~b (G0IV)& $0.359^{+0.022}_{-0.021}$ & $0.654^{+0.060}_{-0.045}$ & $0.04313^{+0.00065}_{-0.00056}$ &  0 & $1.294^{+0.060}_{-0.050}$ & $1.368^{+0.12}_{-0.083}$ & $6.2^{+2.1}_{-0.6}$ &  $-12^{+15}_{-15}$ & $1.9^{+0.9}_{-0.9}$ & TWC08, Wolf07 \\
HD~17156~b (G0) & $3.22^{+0.08}_{-0.08}$ & $1.02^{+0.08}_{-0.08}$ & $0.1614^{+0.0022}_{-0.0022}$ & $0.6801^{+0.0019}_{-0.0019}$ & $1.24^{+0.03}_{-0.03}$ &  $1.44^{+0.08}_{-0.08}$ & $4.18^{+0.31}_{-0.31}$ &  $10.0^{+5.1}_{-5.1}$ & $3.06^{+0.64}_{-0.76}$ & Barbieri09, Narita09, Winn09a  \\
HD~189733~b (K1-2) & $1.138^{+0.022}_{-0.025}$ &  $1.178^{+0.016}_{-0.023}$ & $0.03120^{+0.00027}_{-0.00037}$ & $0.0041^{+0.0025}_{-0.0020}$ & $0.823^{+0.022}_{-0.029}$ & $0.766^{+0.007}_{-0.013}$ & $3.316^{+0.017}_{-0.067}$ &  $-0.85^{+0.32}_{-0.28}$ & $6.8^{+5.2}_{-4.4}$ & Triaud09, TWC08 \\
HD~209458~b (G0V) &  $0.685^{+0.015}_{-0.014}$ & $1.359^{+0.016}_{-0.019}$ & $0.04707^{+0.00046}_{-0.00047}$ & 0 &  $1.119^{+0.033}_{-0.033}$ &  $1.155^{+0.014}_{-0.016}$ & $4.70^{+0.16}_{-0.16}$ &  $-4.4^{+1.4}_{-1.4}$ & $3.1^{+0.8}_{-0.7}$ & TWC08, Winn05 \\
HD~80606~b (G5) & $4.20^{+0.11}_{-0.11}$ & $0.974^{+0.030}_{-0.030}$ & $0.4614^{+0.0047}_{-0.0047}$ & $0.93286^{+0.00055}_{-0.00055}$ & $1.05^{+0.032}_{-0.032}$ &  $0.968^{+0.028}_{-0.028}$ & $1.12^{+0.44}_{-0.22}$ &   $53^{+34}_{-21}$ & $1.6^{+1.8}_{-1.1}$ & Winn09b \\
Kepler-4~b (G0) & $0.077^{+0.012}_{-0.012}$ & $0.357^{+0.019}_{-0.019}$ & $0.0456^{+0.0009}_{-0.0009}$ & 0 (fixed) & $1.223^{+0.053}_{-0.091}$ & $1.487^{+0.071}_{-0.084}$ & $2.2^{+1.0}_{-1.0}$ & & $4.5^{+1.5}_{-1.5}$ & Borucki10 \\
Kepler-5~b (?) & $2.114^{+0.056}_{-0.059}$ & $1.431^{+0.041}_{-0.052}$ & $0.05064^{+0.00070}_{-0.00070}$ & $<0.024$ & $1.374^{+0.040}_{-0.059}$ & $1.793^{+0.043}_{-0.062}$ & $4.8^{+1.0}_{-1.0}$ & & $3.0^{+0.6}_{-0.6}$ & Koch10 \\
Kepler-6~b (F) & $0.669^{+0.025}_{-0.030}$ & $1.323^{+0.026}_{-0.029}$ & $0.04567^{+0.00055}_{-0.00046}$ & 0 (fixed) & $1.209^{+0.044}_{-0.038}$ & $1.391^{+0.017}_{-0.034}$ & $3.0^{+1.0}_{-1.0}$ & & $3.8^{+1.0}_{-1.0}$ & Dunham10 \\
Kepler-7~b (F-G) & $0.433^{+0.040}_{-0.041}$ & $1.478^{+0.050}_{-0.051}$ & $0.06224^{+0.00109}_{-0.00084}$ & 0 (fixed) & $1.347^{+0.072}_{-0.054}$ & $1.843^{+0.048}_{-0.066}$ & $4.2^{+0.5}_{-0.5}$ & & $3.5^{+1.0}_{-1.0}$ & Latham10 \\
Kepler-8~b (F8IV) & $0.603^{+0.13}_{-0.19}$ & $1.419^{+0.056}_{-0.058}$ & $0.0483^{+0.0006}_{-0.0012}$ & 0 (fixed) & $1.213^{+0.067}_{-0.063}$ & $1.486^{+0.053}_{-0.062}$ & $10.5^{+0.7}_{-0.7}$ & $-26.9^{+4.6}_{-4.6}$ & $3.84^{+1.5}_{-1.5}$ & 
Jenkins10 \\
Lupus-TR-3 (K1V) & $0.81^{+0.18}_{-0.18}$ &  $0.89^{+0.07}_{-0.07}$ & $0.0464^{+0.0007}_{-0.0007}$ & 0 (fixed) & $0.87^{+0.04}_{-0.04}$ & $0.82^{+0.05}_{-0.05}$ & & & & Weldrake08 \\
OGLE-TR-10 (G/K) &  $0.62^{+0.14}_{-0.14}$ &  $1.25^{+0.14}_{-0.12}$ &  $0.0434^{+0.0013}_{-0.0015}$ & 0 & $1.14^{+0.10}_{-0.12}$ & $1.17^{+0.13}_{-0.11}$ & $3^{+2}_{-2}$ & & $3.2^{+4.0}_{-3.1}$ & TWC08, Konacki05 \\
OGLE-TR-111 (G/K) & $0.55^{+0.10}_{-0.10}$ & $1.051^{+0.057}_{-0.052}$ & $0.04689^{+0.0010}_{-0.00097}$ & 0 & $0.852^{+0.058}_{-0.052}$ & $0.831^{+0.045}_{-0.040}$ & & & $8.8^{+5.2}_{-6.6}$ & TWC08 \\
OGLE-TR-113 (K) & $1.26^{+0.16}_{-0.16}$ & $1.093^{+0.028}_{-0.019}$ & $0.02289^{+0.00016}_{-0.00015}$ & 0 & $0.779^{+0.017}_{-0.015}$ & $0.774^{+0.020}_{-0.011}$ & & &  $13.2^{+0.8}_{-2.4}$ & TWC08 \\
OGLE-TR-132 (F) & $1.18^{+0.14}_{-0.13}$ & $1.20^{+0.15}_{-0.11}$ & $0.03035^{+0.00057}_{-0.00053}$ & 0 & $1.305^{+0.075}_{-0.067}$ & $1.32^{+0.17}_{-0.12}$ & & & $1.2^{+1.5}_{-1.1}$ & TWC08 \\
OGLE-TR-182 (G) & $1.01^{+0.15}_{-0.15}$ & $1.13^{+0.24}_{-0.08}$ & $0.051^{+0.001}_{-0.001}$ & 0 & $1.14^{+0.05}_{-0.05}$ & $1.14^{+0.23}_{-0.06}$ & & & & Pont08 \\
OGLE-TR-211 (F7-8) & $1.03^{+0.20}_{-0.20}$ & $1.36^{+0.18}_{-0.09}$ & $0.051^{+0.001}_{-0.001}$ & 0 & $1.33^{+0.05}_{-0.05}$ & $1.64^{+0.21}_{-0.07}$ & & & & Udalski08 \\
OGLE-TR-56 (G) &  $1.39^{+0.18}_{-0.17}$ & $1.363^{+0.092}_{-0.090}$ & $0.02383^{+0.00046}_{-0.00051}$ & 0 & $1.228^{+0.072}_{-0.078}$ & $1.363^{+0.089}_{-0.086}$ & $3$ & & $3.2^{+1.0}_{-1.3}$ & TWC08, Konacki03 \\
TrES-1 (K0V) & $0.752^{+0.047}_{-0.046}$ & $1.067^{+0.022}_{-0.021}$ & $0.03925^{+0.00056}_{-0.00060}$ & 0 & $0.878^{+0.038}_{-0.040}$ & $0.807^{+0.017}_{-0.016}$ & $1.3^{+0.3}_{-0.3}$ & $30^{+21}_{-21}$ & $3.7^{+3.4}_{-2.8}$ & TWC08, Narita07 \\
TrES-2 (G0V) & $1.200^{+0.051}_{-0.053}$ & $1.224^{+0.041}_{-0.041}$ & $0.03558^{+0.00070}_{-0.00077}$ &  0 & $0.983^{+0.059}_{-0.063}$ & $1.003^{+0.033}_{-0.033}$& $1.0^{+0.6}_{-0.6}$ & $-9.0^{+12.0}_{-12.0}$ & $5.0^{+2.7}_{-2.1}$ & TWC08, Winn08 \\
TrES-3 (G) & $1.938^{+0.062}_{-0.063}$ & $1.312^{+0.033}_{-0.041}$ & $0.02272^{+0.00017}_{-0.00026}$ & 0 & $0.915^{+0.021}_{-0.031}$ & $0.812^{+0.014}_{-0.025}$ & $1.5^{+1.0}_{-1.0}$ & & $0.6^{+2.0}_{-0.4}$ & TWC08 \\
TrES-4 (F) & $0.920^{+0.073}_{-0.072}$ & $1.751^{+0.064}_{-0.062}$ & $0.05092^{+0.00072}_{-0.00069}$ & 0 & $1.394^{+0.060}_{-0.056}$ & $1.816^{+0.065}_{-0.062}$ & $8.5^{+1.2}_{-1.2}$ & $6.3^{+4.7}_{-4.7}$ & $2.9^{+0.4}_{-0.4}$ & TWC08, Narita10 \\
WASP-1~b (F7V) & $0.918^{+0.091}_{-0.090}$ & $1.514^{+0.052}_{-0.047}$ & $0.03957^{+0.00049}_{-0.00048}$ & 0 & $1.301^{+0.049}_{-0.047}$ & $1.517^{+0.052}_{-0.045}$ & $5.79^{+0.35}_{-0.35}$ & & $3.0^{+0.6}_{-0.6}$ & TWC08, Stempels07 \\ 
WASP-10~b (K5) & $2.96^{+0.22}_{-0.17}$ & $1.28^{+0.077}_{-0.091}$ & $0.0369^{+0.0012}_{-0.0014}$ & $0.059^{+0.014}_{-0.004}$ & $0.703^{+0.068}_{-0.080}$ & $0.775^{+0.043}_{-0.040}$ & $<6$ & & $0.8^{+0.2}_{-0.2}$ & Christian09 \\
WASP-11~b (K3V) & $0.487^{+0.018}_{-0.018}$ & $1.005^{+0.032}_{-0.027}$ & $0.0435^{+0.0006}_{-0.0006}$ & 0 (fixed) & $0.83^{+0.03}_{-0.03}$ &  $0.79^{+0.02}_{-0.02}$ &  $0.5^{+0.2}_{-0.2}$ & & $7.9^{+3.8}_{-3.8}$ & Bakos09b \\
WASP-12~b (G0) & $1.41^{+0.10}_{-0.10}$ & $1.79^{+0.09}_{-0.09}$ & $0.0229^{+0.0008}_{-0.0008}$ & $0.049^{+0.015}_{-0.015}$ & $1.35^{+0.14}_{-0.14}$ & $1.57^{+0.07}_{-0.07}$ & $<2.2^{+1.5}_{-1.5}$ & & $2^{+1}_{-1}$ & Hebb09 \\
WASP-13~b (G1V) & $0.46^{+0.06}_{-0.05}$ & $1.21^{+0.14}_{-0.12}$ & $0.0527^{+0.0017}_{-0.0019}$ & 0 (fixed) & $1.03^{+0.11}_{-0.09}$ & $1.34^{+0.13}_{-0.11}$ & $<4.9$ & & $8.5^{+5.5}_{-4.9}$ & Skillen09 \\
WASP-14~b (F5V) & $7.341^{+0.508}_{-0.496}$ & $1.281^{+0.075}_{-0.082}$ & $0.036^{+0.001}_{-0.001}$ & $0.091^{+0.003}_{-0.003}$ & $1.211^{+0.127}_{-0.122}$ & $1.306^{+0.066}_{-0.073}$ & $4.9^{+1.0}_{-1.0}$ & $-33.1^{+7.4}_{-7.4}$ & $\sim$0.5-1.0 & Joshi09, Johnson09 \\
WASP-15~b (F5) & $0.542^{+0.050}_{-0.050}$ & $1.428^{+0.077}_{-0.077}$ & $0.0499^{+0.0018}_{-0.0018}$ & 0 (fixed) & $1.18^{+0.12}_{-0.12}$ & $1.477^{+0.072}_{-0.072}$ & $4.27^{+0.26}_{-0.36}$ & $-139.6^{+4.3}_{-5.2}$ & $3.9^{+2.8}_{-1.3}$ & 
West09, Triaud10 \\
WASP-16~b (G3V) & $0.855^{+0.043}_{-0.076}$ & $1.008^{+0.083}_{-0.060}$ & $0.0421^{+0.0010}_{-0.0018}$ & 0 (fixed) & $1.022^{+0.074}_{-0.129}$ & $0.946^{+0.057}_{-0.052}$ & $3.0^{+1.0}_{-1.0}$ & & $2.3^{+5.8}_{-2.2}$ & Lister09 \\
WASP-17~b (F6) & $0.490^{+0.059}_{-0.056}$ & $1.74^{+0.26}_{-0.23}$ & $0.0501^{+0.0017}_{-0.0018}$ & $0.129^{+0.106}_{-0.068}$ & $1.20^{+0.12}_{-0.12}$ & $1.38^{+0.20}_{-0.18}$ & $10.14^{+0.58}_{-0.79}$ & $-147.3^{+5.5}_{-5.9}$ & $3.0^{+0.9}_{-2.6}$ & Anderson10, Triaud10 \\
WASP-18~b (F9) & $10.30^{+0.69}_{-0.69}$ & $1.106^{+0.072}_{-0.054}$ & $0.02026^{+0.00068}_{-0.00068}$ & $0.0092^{+0.0028}_{-0.0028}$ & $1.25^{+0.13}_{-0.13}$ & $1.216^{+0.067}_{-0.054}$ & $14.67^{+0.81}_{-0.57}$ & $5.0^{+2.8}_{-3.1}$ & 0.5-1.5 &
Hellier09a, Triaud10 \\
WASP-19~b (G8V) & $1.14^{+0.07}_{-0.07}$ & $1.28^{+0.07}_{-0.07}$ & $0.0164^{+0.0005}_{-0.0006}$ & $0.02^{+0.02}_{-0.01}$ & $0.95^{+0.10}_{-0.10}$ & $0.93^{+0.05}_{-0.04}$ & $4^{+2}_{-2}$ & & $\gtrsim 1$ & Hebb10 \\
WASP-2~b (K1V) &  $0.915^{+0.090}_{-0.093}$ & $1.071^{+0.080}_{-0.083}$ & $0.03138^{+0.00130}_{-0.00154}$ & 0 & $0.89^{+0.12}_{-0.12}$ &  $0.840^{+0.062}_{-0.065}$ & $0.99^{+0.27}_{-0.32}$ & $-153^{+15}_{-11}$ & $5.6^{+8.4}_{-5.6}$ & 
TWC08, Triaud10 \\ 
WASP-3~b (F7V) &  $1.76^{+0.08}_{-0.14}$ & $1.31^{+0.07}_{-0.14}$ & $0.0317^{+0.0005}_{-0.0010}$ & 0 & $1.24^{+0.06}_{-0.11}$ &  $1.31^{+0.06}_{-0.12}$ & $13.4^{+1.5}_{-1.5}$ & $15^{+10}_{-9}$ & 0.7-3.5 & Pollacco08, Simpson09 \\ 
WASP-4~b (G7V) &  $1.21^{+0.13}_{-0.08}$ & $1.304^{+0.054}_{-0.042}$ & $0.02255^{+0.00095}_{-0.00065}$ & 0 (fixed) & $0.85^{+0.11}_{-0.07}$ & $0.873^{+0.036}_{-0.027}$ & $2.14^{+0.38}_{-0.35}$ & $4^{+34}_{-43}$ & $5.2^{+3.8}_{-3.2}$ & 
Gillon09b, Triaud10 \\
WASP-5~b (G4V) &  $1.58^{+0.13}_{-0.10}$ & $1.087^{+0.068}_{-0.071}$ & $0.0267^{+0.0012}_{-0.0008}$ & $0.038^{+0.026}_{-0.018}$ & $0.96^{+0.13}_{-0.09}$ & $1.029^{+0.056}_{-0.069}$ & $3.24^{+0.34}_{-0.35}$ & $12.4^{+8.2}_{-11.9}$ & $5.4^{+4.4}_{-4.3}$ & Gillon09b, Triaud10 \\
WASP-6~b (G8V) &  $0.503^{+0.019}_{-0.038}$ & $1.224^{+0.051}_{-0.052}$ & $0.0421^{+0.0008}_{-0.0013}$ & $0.054^{+0.018}_{-0.015}$ & $0.880^{+0.050}_{-0.080}$ & $0.870^{+0.025}_{-0.036}$ & $1.4^{+1.0}_{-1.0}$ & $0.20^{+0.25}_{-0.32}$ & $11^{+7}_{-7}$ & Gillon09a \\
WASP-7~b (F5V) &  $0.96^{+0.12}_{-0.18}$ & $0.915^{+0.046}_{-0.040}$ & $0.0618^{+0.0014}_{-0.0033}$ & 0 (fixed) & $1.28^{+0.09}_{-0.19}$ &  $1.236^{+0.059}_{-0.046}$ & $17^{+2}_{-2}$ & & & Hellier09b \\ 
XO-1~b (G1V) & $0.918^{+0.081}_{-0.078}$ & $1.206^{+0.047}_{-0.042}$ & $0.04928^{+0.00089}_{-0.00099}$ & 0 & $1.027^{+0.057}_{-0.061}$ & $0.934^{+0.037}_{-0.032}$ & $1.11^{+0.67}_{-0.67}$ &  &  $1.0^{+3.1}_{-0.9}$  & 
TWC08, McCullough06 \\ 
XO-2~b (K0V) & $0.566^{+0.055}_{-0.055}$ & $0.983^{+0.029}_{-0.028}$ & $0.03684^{0.00040}_{-0.00043}$ & 0 & $0.974^{+0.032}_{-0.034}$ & $0.971^{+0.027}_{-0.026}$ & $1.4^{+0.3}_{-0.3}$ & &  $5.8^{+2.8}_{-2.3}$ & TWC08, Burke07 \\
XO-3~b (F5V) & $13.25^{+0.64}_{-0.64}$ & $1.95^{+0.16}_{-0.16}$ & $0.0476^{+0.0005}_{-0.0005}$ & $0.260^{+0.017}_{-0.017}$ &  $1.41^{+0.03}_{-0.05}$ & $2.13^{+0.04}_{-0.05}$ & $18.54^{+0.17}_{-0.17}$ &  $37.3^{+3.7}_{-3.7}$ &  $2.69^{+0.14}_{-0.16}$ & Johns-Krull08, Winn09d \\ 
XO-4~b (F5V) & $1.72^{+0.20}_{-0.20}$ & $1.34^{+0.048}_{-0.048}$ & $0.0555^{+0.0011}_{-0.0011}$ & 0 (fixed) & $1.32^{+0.02}_{-0.02}$ & $1.56^{+0.05}_{-0.05}$ &  $8.8^{+0.5}_{-0.5}$ & & $2.1^{+0.6}_{-0.6}$ & McCullough08 \\ 
XO-5~b (G8V) & $1.059^{+0.028}_{-0.028}$ & $1.109^{+0.050}_{-0.050}$ & $0.0488^{+0.0006}_{-0.0006}$ &  0 & $0.88^{+0.03}_{-0.03}$ & $1.08^{+0.04}_{-0.04}$ &  $0.7^{+0.5}_{-0.5}$ & & $14.8^{+2.0}_{-2.0}$ & P\'{a}l09 \\ 
\enddata

\tablecomments{Column 1 -- planet's name and the stellar spectral type inside the bracket, Column 2 -- planetary mass, Column 3 -- planetary radius, Column 4 -- semi-major axis, Column 5 -- eccentricity, Column 6 -- stellar mass, Column 7 -- stellar radius, Column 8 -- projected stellar rotational velocity, Column 9 -- projected stellar obliquity, Column 10 -- stellar age, Column 11 -- references, where we indicate the references by the first author name and the published year as, for example, Barge08 for \cite{Barge08}.  TWH08 is \cite{Torres08}}

\end{deluxetable}
\clearpage
\end{landscape}
%
%
\begin{deluxetable}{lccccccc}

\tabletypesize{\footnotesize}
\tablecaption{Critical conditions for tidal instability, as well as the Roche limit for each system in Table~\ref{tab1}. \label{tab2}}
\tablewidth{500pt}
\tablecolumns{8} \tablehead{ \colhead{Planet Name} &
\colhead{$n/\omega_*$} & \colhead{$n/(\omega_*\cos \epsilon)$} &
\colhead{$P_{rot,*}\,$[days]} &  \colhead{$L_{\rm tot}/L_c$} &
\colhead{$L_{\rm *,spin}/L_{\rm orb}$} & \colhead{$(1/3-L_{\rm *,spin}/L_{\rm orb})$} &
\colhead{$a/a_R$} }

\startdata
CoRoT-1 & $ 7.121$ & $31.657$ & $10.796$ & $ 0.612$ & $ 0.336$ &  & $ 1.667$ \\
CoRoT-2 & $ 2.208$ & $ 2.225$ & $ 3.850$ & $ 0.844$ & $ 0.186$ &  & $ 2.749$ \\
CoRoT-3 & $ 0.515$ & $ 0.650$ & $ 2.176$ & $ 1.272$ & $ 0.125$ & $ 0.208$ & $13.639$ \\
CoRoT-4 & $ 1.010$ & $ 1.010$ & $ 9.246$ & $ 0.998$ & $ 0.366$ &  & $ 6.141$ \\
CoRoT-5 & $14.930$ & $14.930$ & $59.986$ & $ 0.555$ & $ 0.113$ &  & $ 2.632$ \\
CoRoT-6 & $ 0.767$ & $ 0.767$ & $ 6.821$ & $ 1.205$ & $ 0.091$ & $ 0.242$ & $ 9.860$ \\
CoRoT-7 & $ 0.000$ & $ 0.000$ & $ 0.000$ & $ 0.149$ & $ 0.000$ &  & $ 2.764$ \\
GJ1214 & $ 3.385$ & $ 3.385$ & $ 5.336$ & $ 0.759$ & $ 0.698$ &  & $ 2.885$ \\
GJ436 & $17.070$ & $17.070$ & $45.131$ & $ 0.524$ & $ 0.130$ &  & $ 3.950$ \\
HAT-P-1 & $ 3.432$ & $ 3.439$ & $15.308$ & $ 0.734$ & $ 0.356$ &  & $ 3.294$ \\
HAT-P-11 & $ 5.108$ & $ 5.108$ & $25.289$ & $ 0.671$ & $ 0.543$ &  & $ 5.549$ \\
HAT-P-12 & $22.093$ & $22.093$ & $70.911$ & $ 0.545$ & $ 0.071$ &  & $ 2.517$ \\
HAT-P-13 & $ 9.328$ & $ 9.328$ & $27.208$ & $ 0.574$ & $ 0.278$ &  & $ 2.816$ \\
HAT-P-2 & $ 0.708$ & $ 0.708$ & $ 3.988$ & $ 0.995$ & $ 0.192$ &  & $10.659$ \\
HAT-P-3 & $29.067$ & $29.067$ & $84.263$ & $ 0.593$ & $ 0.034$ &  & $ 3.546$ \\
HAT-P-4 & $ 4.772$ & $ 4.772$ & $14.622$ & $ 0.712$ & $ 0.671$ &  & $ 2.722$ \\
HAT-P-5 & $ 8.142$ & $ 8.142$ & $22.702$ & $ 0.623$ & $ 0.150$ &  & $ 2.987$ \\
HAT-P-6 & $ 2.205$ & $ 2.205$ & $ 8.488$ & $ 0.848$ & $ 0.585$ &  & $ 3.506$ \\
HAT-P-7 & $11.113$ & $-11.124$ & $24.490$ & $ 0.553$ & $ 0.241$ &  & $ 2.804$ \\
HAT-P-8 & $ 2.004$ & $ 2.004$ & $ 6.949$ & $ 0.871$ & $ 0.602$ &  & $ 3.273$ \\
HAT-P-9 & $ 1.425$ & $ 1.425$ & $ 5.610$ & $ 1.036$ & $ 0.971$ & $-0.638$ & $ 3.055$ \\
HD149026 & $ 3.881$ & $ 3.968$ & $11.160$ & $ 0.868$ & $ 1.270$ &  & $ 4.094$ \\
HD17156 & $ 2.286$ & $ 2.294$ & $48.555$ & $ 0.945$ & $ 0.025$ &  & $20.702$ \\
HD189733 & $ 5.269$ & $ 5.270$ & $11.684$ & $ 0.722$ & $ 0.113$ &  & $ 2.809$ \\
HD209458 & $ 3.526$ & $ 3.537$ & $12.429$ & $ 0.731$ & $ 0.379$ &  & $ 2.799$ \\
HD80606 & $ 0.392$ & $ 0.651$ & $43.714$ & $ 1.054$ & $ 0.011$ & $ 0.323$ & $71.577$ \\
Kepler-4 & $10.631$ & $10.631$ & $34.186$ & $ 0.822$ & $ 2.159$ &  & $ 4.837$ \\
Kepler-5 & $ 5.325$ & $ 5.325$ & $18.893$ & $ 0.671$ & $ 0.208$ &  & $ 3.889$ \\
Kepler-6 & $ 7.236$ & $ 7.236$ & $23.451$ & $ 0.610$ & $ 0.315$ &  & $ 2.698$ \\
Kepler-7 & $ 4.543$ & $ 4.543$ & $22.194$ & $ 0.746$ & $ 0.816$ &  & $ 2.746$ \\
Kepler-8 & $ 2.034$ & $ 2.281$ & $ 7.158$ & $ 1.021$ & $ 1.273$ & $-0.939$ & $ 2.567$ \\
OGLE-TR-10 & $ 6.380$ & $ 6.380$ & $19.725$ & $ 0.631$ & $ 0.285$ &  & $ 2.698$ \\
OGLE-TR-111 & $ 0.000$ & $ 0.000$ & $ 0.000$ & $ 0.632$ & $ 0.000$ & & $ 3.670$ \\
OGLE-TR-113 & $ 0.000$ & $ 0.000$ & $ 0.000$ & $ 0.575$ & $ 0.000$ &  & $ 2.340$ \\
OGLE-TR-132 & $ 0.000$ & $ 0.000$ & $ 0.000$ & $ 0.439$ & $ 0.000$ &  & $ 2.328$ \\
OGLE-TR-56 & $18.964$ & $18.964$ & $22.979$ & $ 0.489$ & $ 0.207$ &  & $ 1.734$ \\
TrES-1 & $10.363$ & $11.966$ & $31.397$ & $ 0.670$ & $ 0.065$ &  & $ 3.325$ \\
TrES-2 & $20.531$ & $20.787$ & $50.730$ & $ 0.613$ & $ 0.043$ &  & $ 2.957$ \\
TrES-3 & $20.960$ & $20.960$ & $27.380$ & $ 0.622$ & $ 0.039$ &  & $ 2.117$ \\
TrES-4 & $ 3.041$ & $ 3.060$ & $10.806$ & $ 0.834$ & $ 0.862$ &  & $ 2.410$ \\
WASP-1 & $ 5.259$ & $ 5.259$ & $13.252$ & $ 0.676$ & $ 0.538$ &  & $ 2.215$ \\
WASP-10 & $ 2.120$ & $ 2.120$ & $ 6.533$ & $ 0.988$ & $ 0.068$ &  & $ 4.431$ \\
WASP-11 & $21.978$ & $21.978$ & $79.914$ & $ 0.631$ & $ 0.035$ &  & $ 3.449$ \\
WASP-12 & $33.152$ & $33.152$ & $36.094$ & $ 0.429$ & $ 0.185$ &  & $ 1.236$ \\
WASP-13 & $ 3.178$ & $ 3.178$ & $13.832$ & $ 0.779$ & $ 0.619$ &  & $ 3.169$ \\
WASP-14 & $ 5.964$ & $ 7.119$ & $13.481$ & $ 0.807$ & $ 0.050$ &  & $ 4.877$ \\
WASP-15 & $ 4.984$ & $ 4.984$ & $18.676$ & $ 0.683$ & $ 0.520$ &  & $ 2.566$ \\
WASP-16 & $ 5.113$ & $ 5.113$ & $15.949$ & $ 0.695$ & $ 0.161$ &  & $ 3.746$ \\
WASP-17 & $ 2.075$ & $-2.474$ & $ 7.755$ & $ 0.999$ & $ 1.228$ & $-0.894$ & $ 2.033$ \\
WASP-18 & $ 5.958$ & $ 5.958$ & $ 5.591$ & $ 0.713$ & $ 0.100$ &  & $ 3.522$ \\
WASP-19 & $14.951$ & $14.951$ & $11.759$ & $ 0.513$ & $ 0.244$ &  & $ 1.296$ \\
WASP-2 & $ 0.000$ & $ 0.000$ & $ 0.000$ & $ 0.578$ & $ 0.000$ &  & $ 2.815$ \\
WASP-3 & $ 2.673$ & $ 2.767$ & $ 4.945$ & $ 0.812$ & $ 0.612$ &  & $ 2.589$ \\
WASP-4 & $16.469$ & $16.469$ & $22.077$ & $ 0.566$ & $ 0.087$ &  & $ 1.852$ \\
WASP-5 & $ 9.151$ & $ 9.151$ & $14.870$ & $ 0.614$ & $ 0.134$ &  & $ 2.760$ \\
WASP-6 & $ 9.348$ & $ 9.348$ & $31.431$ & $ 0.629$ & $ 0.109$ &  & $ 2.717$ \\
WASP-7 & $ 0.742$ & $ 0.742$ & $ 3.677$ & $ 1.221$ & $ 0.978$ & $-0.644$ & $ 5.841$ \\
XO-1 & $10.799$ & $10.799$ & $42.559$ & $ 0.696$ & $ 0.051$ &  & $ 3.747$ \\
XO-2 & $13.409$ & $13.409$ & $35.080$ & $ 0.566$ & $ 0.121$ &  & $ 2.977$ \\
XO-3 & $ 1.827$ & $ 2.297$ & $ 5.811$ & $ 0.872$ & $ 0.166$ &  & $ 4.903$ \\
XO-4 & $ 2.159$ & $ 2.159$ & $ 8.966$ & $ 0.826$ & $ 0.382$ &  & $ 4.306$ \\
XO-5 & $18.603$ & $18.603$ & $78.035$ & $ 0.680$ & $ 0.030$ &  & $ 4.455$ \\
\enddata

\end{deluxetable}

\end{document}